\documentclass[sigconf]{acmart}

\usepackage{etoolbox}
\usepackage{enumitem}

\usepackage{comment}
\usepackage{xspace}
\usepackage{xr-hyper}
\usepackage{url}
\usepackage{graphicx}
\usepackage{balance}  
\usepackage{xcolor}
\usepackage{amsmath}
\usepackage{enumitem}
\usepackage{subfigure}
\usepackage{cleveref}
\usepackage{url}
\usepackage{multirow}
\usepackage{amsmath}
\usepackage{float}
\usepackage{framed}
\usepackage{xcolor} 

\definecolor{shadecolor}{rgb}{0.9,0.9,0.9} 

\usepackage[ruled,linesnumbered]{algorithm2e}

\newcommand{\stitle}[1]{\vspace{1mm}\noindent{\bf #1}}

\newtheorem{example}{Example}
\makeatletter
\DeclareRobustCommand*\cal{\@fontswitch\relax\mathcal}
\makeatother

\newcommand{\sys}{\textsc{Andromeda}\xspace}

\newcommand{\etitle}[1]{\vspace{1mm}\noindent{\underline{\em #1}}}
\newcommand{\revised}[1]{{\color{black}#1}}

\AtBeginDocument{%
  \providecommand\BibTeX{{%
    \normalfont B\kern-0.5em{\scshape i\kern-0.25em b}\kern-0.8em\TeX}}}

\setcopyright{acmlicensed}
\acmJournal{PACMMOD}
\acmYear{2025} \acmVolume{3} \acmNumber{N1 (SIGMOD)}
\acmArticle{013} \acmMonth{1} \acmPrice{15.00}
\acmDOI{10.1145/3709663}

\received{July 2024}
\received[revised]{September 2024}
\received[accepted]{November 2024}

\begin{document}

\newcommand{\ie}{{\em i.e.,}\xspace}
\newcommand{\eg}{{\em e.g.,}\xspace}
\newcommand{\wrt}{\emph{w.r.t.}\xspace}
\newcommand{\aka}{\emph{a.k.a.}\xspace}
\newcommand{\kwlog}{\emph{w.l.o.g.}\xspace}
\newcommand{\Kset}{\mathbb{K}\xspace}
\newcommand{\Dset}{\mathbb{D}\xspace}
\newcommand{\Qset}{{Q}\xspace}
\newcommand{\Mset}{{M}\xspace}
\newcommand{\Sset}{{S}\xspace}
\newcommand{\qhis}{q^{\tt H}\xspace}
\newcommand{\mhis}{m\xspace}
\newcommand{\shis}{s\xspace}
\newcommand{\enc}{f_{\tt enc}\xspace}

\newcommand{\term}[1]{{\tt #1}}


\title{Automatic Database Configuration Debugging using Retrieval-Augmented Language Models}




\author{Sibei Chen}
\affiliation{%
  \institution{Renmin University of China}
  }
\orcid{0009-0001-5331-5829}
\email{sibei@ruc.edu.cn}

\author{Ju Fan}
\thanks{$^*$ Ju Fan is the corresponding author.}
\affiliation{%
  \institution{Renmin University of China}
  }
\orcid{0000-0003-4729-9903}
\email{fanj@ruc.edu.cn}

\author{Bin Wu}
\affiliation{%
  \institution{Alibaba Cloud Computing}
  }
\orcid{0000-0002-4743-1006}
\email{binwu.wb@alibaba-inc.com}

\author{Nan Tang}
\affiliation{%
  \institution{HKUST (GZ) / HKUST}
  }
\orcid{0000-0003-2832-0295}
\email{nantang@hkust-gz.edu.cn}

\author{Chao Deng}
\affiliation{%
  \institution{Renmin University of China}
  }
\orcid{0009-0003-9312-2110}
\email{dengc@ruc.edu.cn}

\author{Pengyi Wang}
\affiliation{%
  \institution{Renmin University of China}
  }
\orcid{0009-0009-5634-9782}
\email{2023103709@ruc.edu.cn}

\author{Ye Li}
\affiliation{%
  \institution{Alibaba Cloud Computing}
  }
\orcid{0000-0003-1860-2723}
\email{liye.li@alibaba-inc.com}

\author{Jian Tan}
\affiliation{%
  \institution{Alibaba Cloud Computing}
  \country{The United States}
  }
\orcid{0000-0002-1080-9300}
\email{j.tan@alibaba-inc.com}

\author{Feifei Li}
\affiliation{%
  \institution{Alibaba Cloud Computing}
  }
\orcid{0009-0003-0770-5775}
\email{lifeifei@alibaba-inc.com}

\author{Jingren Zhou}
\affiliation{%
  \institution{Alibaba Cloud Computing}
  }
\orcid{0000-0002-4220-2634}
\email{jingren.zhou@alibaba-inc.com}

\author{Xiaoyong Du}
\affiliation{%
  \institution{Renmin University of China}
  }
\orcid{0000-0002-5757-9135}
\email{duyong@ruc.edu.cn}





\renewcommand{\shortauthors}{Sibei Chen et al.}

\begin{abstract}

Database management system (DBMS) configuration debugging, \eg diagnosing poorly configured DBMS knobs and generating troubleshooting recommendations, is crucial in optimizing DBMS performance.
However,  the configuration debugging process is tedious and, sometimes challenging, even for seasoned database administrators (DBAs) with sufficient experience in DBMS configurations and good understandings of the DBMS internals (\eg MySQL or Oracle).
To address this difficulty, we propose \sys, a framework that utilizes large language models (LLMs) to enable automatic DBMS configuration debugging. \sys serves as a natural surrogate of DBAs to answer a wide range of natural language (NL) questions on DBMS configuration issues, and to generate diagnostic suggestions to fix these issues. Nevertheless, directly prompting LLMs with these professional questions may result in overly generic and often unsatisfying answers. To this end, we propose a retrieval-augmented generation (RAG) strategy that effectively provides matched \emph{domain-specific contexts} for the question from multiple sources. They come from related historical questions, troubleshooting manuals and DBMS telemetries, which significantly improve the performance of configuration debugging. To support the RAG strategy, we develop a document retrieval mechanism addressing heterogeneous documents and design an effective method for telemetry analysis. Extensive experiments on real-world DBMS configuration debugging datasets show that \sys significantly outperforms existing solutions.
\end{abstract}


\maketitle

\section{Introduction}
Database management system (DBMS) performance diagnosis is a recurring theme, which constantly baffles cloud database customers.  
Poorly configured DBMSs may suffer from unexpected performance pains. 
There has been a growing focus on \emph{automatic DBMS configuration debugging}. 
Traditional approaches~\cite{DBLP:conf/cidr/DiasRSVW05,DBLP:conf/sigmod/YoonNM16,tuneful,ottertune, iTuned, ResTune, QTune} focus on building complex models from performance metrics (such as runtime, resource usages, etc.) to analyze the root cause of a performance issue, and then identifying potential configurations (\eg problematic knobs) to tune. 

\begin{figure*}[t]
	\centering 
	\includegraphics[width=\textwidth]{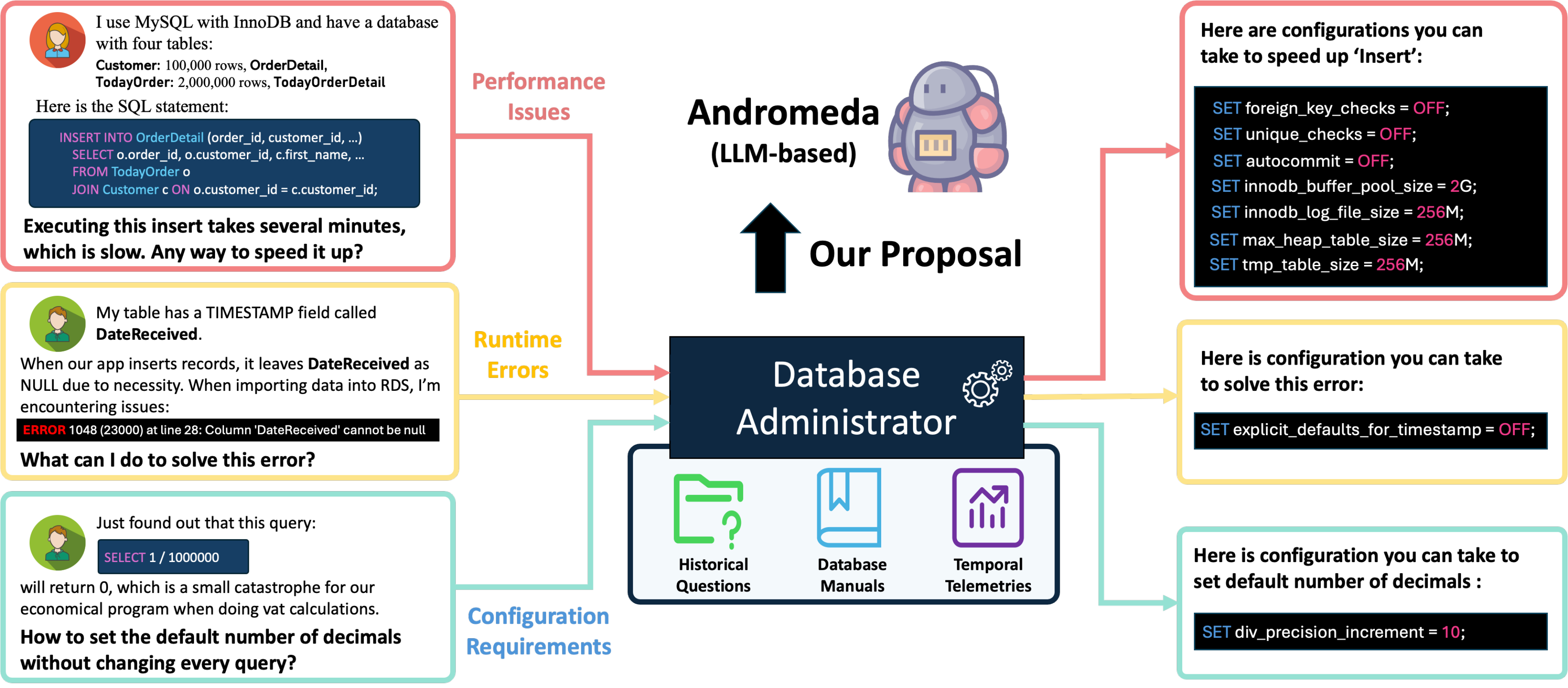}
	\caption{Overview of automatic DBMS configuration debugging, where users directly pose NL debugging questions regarding configuration issues, and there exists a ``co-pilot'' to diagnose the issues and generate recommendations to fix the issues.}
	\label{fig:overview} 
\end{figure*}

\stitle{Human-based Debugging.}
In real-world scenarios, users often pose natural language (NL) questions regarding configuration issues. Database administrators (DBAs) then perform configuration debugging, as illustrated in Figure~\ref{fig:overview}.
Specifically, database users can query in NL on \emph{diverse} configuration debugging requirements, including diagnosing performance issues (\eg ``\emph{Executing an INSERT statement is very slow}''), execution errors (\eg ``\emph{Importing data into RDS encounters issues}'') and sub-optimal configurations (\eg ``The number of decimals does not meet my application requirements''), or even requesting for troubleshooting recommendations (\eg ``\emph{How to fix the issue}'' and ``\emph{How to specify proper configurations}'').
Given such NL debugging questions, a DBA typically analyzes information from multiple sources, such as related historical questions, troubleshooting manuals and DBMS telemetry data (\ie performance metrics, like CPU utilization and query latency).
Based on this analysis, the DBA provides recommendations on configuration settings to resolve the issues (as shown in the right part of Figure~\ref{fig:overview}).

Clearly, the above process is challenging and time-consuming, 
as it requires the DBA to be highly experienced in configuration debugging and to have a comprehensive understanding of the DBMS's internal working mechanisms. 
This raises a crucial question: \emph{Can we replace the DBA with an LLM-based agent to enable automatic DBMS configuration debugging?}

\stitle{Our Proposal: LLM-based Debugging.}
To address the aforementioned problem, we propose \sys, a framework that harnesses large language models (LLMs) to support automatic DBMS configuration debugging. 
\sys serves as a natural surrogate of DBAs to answer a wide range of NL questions regarding DBMS configuration issues, as shown in Figure~\ref{fig:overview}. A straightforward approach to support \sys is to directly prompt an existing LLM (\eg GPT-4) with NL debugging questions and return the answer of the LLM.
However, this approach often yields \emph{overly-generic yet useless} recommendations.
This limitation stems from the fact that LLMs are trained on general-purpose datasets and lack the specific domain knowledge related to DBMS configuration.
To solve this problem, \sys employs a retrieval-augmented generation (RAG) strategy that 
enriches the NL debugging questions with domain-specific context drawn from multiple sources,
including historical questions, troubleshooting manuals and DBMS telemetry data. 
We further illustrate the strategy in the following example. 

\begin{figure*}
	\centering
	%
	\centering   
	\includegraphics[width=\textwidth]{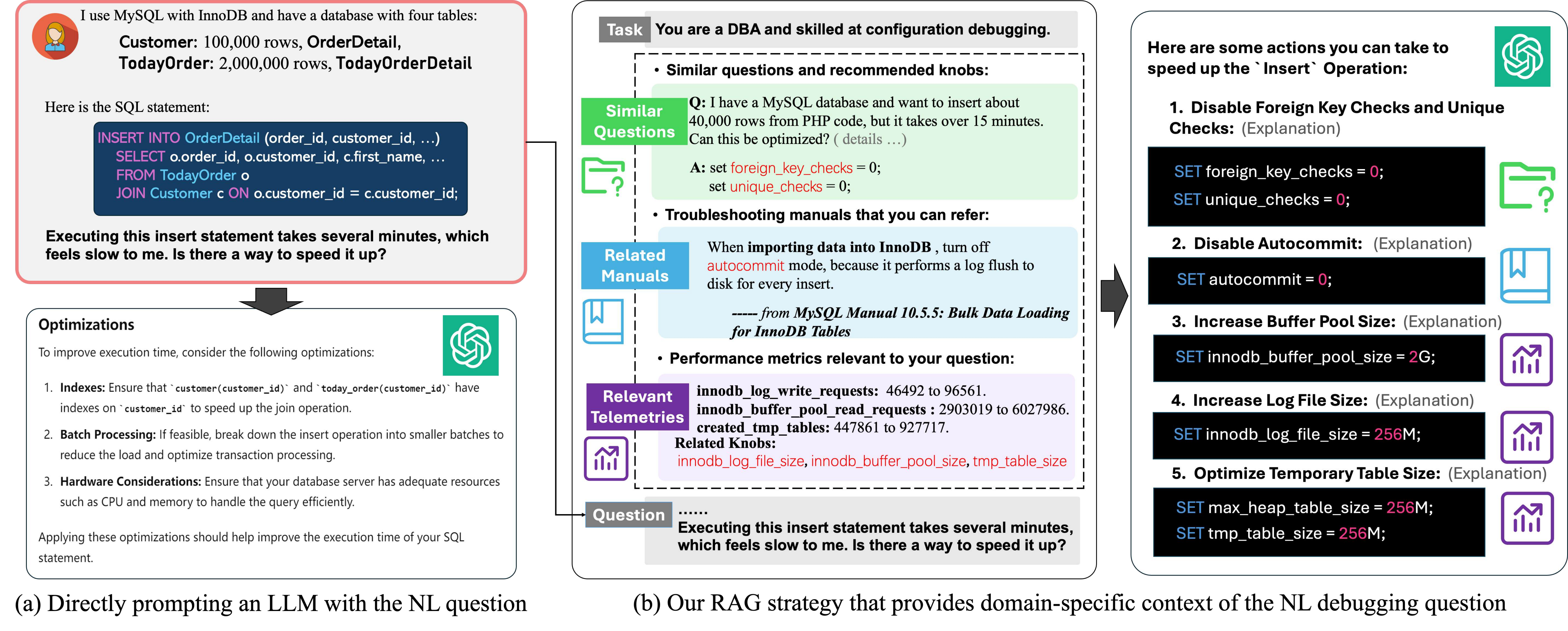}
	\caption{An example of our RAG strategy in \sys. 
	(a) A straightforward strategy that directly prompts an LLM with the NL question results in overly-generic yet useless answers. 
	(b) Our RAG strategy provides \emph{domain-specific context} of an NL debugging question from multiple sources 
	that improve the inference capabilities of the LLM on configuration debugging.}
	\label{fig:running-example}
\end{figure*}

\begin{example}
Let's consider an NL question regarding the slow execution of an INSERT statement, as shown in Figure~\ref{fig:running-example}.
We can see that the results of directly prompting an existing LLM (\eg GPT-4) are vague and overly generic, although they are ``technically correct'' (Figure~\ref{fig:running-example} (a)), which are unhelpful for the users to solve the issue. 
In contrast, as shown in Figure~\ref{fig:running-example} (b), equipped with our RAG strategy, \sys finds contextual information from the following sources: 

(a) \sys finds similar historical debugging questions that another user also encounters slow execution during data insertion, and provides a corresponding configuration solution, \ie disabling \texttt{foreign\_key\_checks} and \texttt{unique\_checks}, for reference. 
	
(b) \sys identifies a document from the \textit{MySQL Manual 10.5.5: Bulk Data Loading for InnoDB Tables}, which describes that disabling \texttt{autocommit} when importing large amounts of data can achieve speed-up and explains the reasons.
	
(c) \sys considers the telemetry log and detects three troublesome performance metrics, \eg \texttt{innodb\_log\_write\_requests}, and suggests related knobs to tune. 

With the above contextual information, \sys leverages LLMs to diagnose improper DBMS configurations and generates accurate troubleshooting recommendations.

\end{example}

\stitle{Key Technical Challenges.}

\etitle{(C1) Retrieving heterogeneous documents.}
We retrieve documents from multiple sources, such as historical questions and DBMS manuals, where the semantics of documents from different sources may be heterogeneous. For example, as shown in Figure~\ref{fig:running-example}, historical questions typically describe ``\emph{what is wrong}'' (\eg ``\emph{It takes over 15 minutes $\ldots$}''), while DBMS manuals focus on ``\emph{How to fix it}'' (\eg ``\emph{turn off \texttt{autocommit} mode}''). 
This heterogeneity poses significant challenges on retrieval mechanisms. 

\etitle{(C2) Detecting troublesome telemetries.}
We also extract relevant telemetry data, \ie time series data of a performance metric (such as CPU utilization) with timestamps,  which is related to the user's NL question. However, there are many types of telemetries in a DBMS, and only a small number of telemetries are relevant to the question. Moreover, even given a relevant telemetry, only a small proportion of timestamps (\eg the duration of high CPU utilization) are useful. Thus, it is non-trivial to detect such troublesome telemetries that are relevant to a user's NL question. 




To address challenge (C1), we propose an effective document retrieval mechanism that addresses the semantic heterogeneity among different sources. We propose a contrastive learning approach that aligns documents from different sources into a unified representation space, and enables accurate document retrieval on the space. We also design a novel data synthesis method to tackle the scarcity of training data. 
To address challenge (C2), we introduce a telemetry detection method to identify troublesome telemetries that are relevant to a user's NL question. The method first utilizes a seasonal-trend decomposition mechanism to find troublesome telemetries, and then judiciously selects relevant telemetries. 

\stitle{Contributions.} Our contributions are summarized as follows.

\vspace{1mm} \noindent
(1) {\bf LLM-based automatic DBMS configuration debugging framework.}
We formally define the problem of automatic DBMS configuration debugging (Section~\ref{sec:pre}) and introduce an overview of the \sys framework (Section~\ref{sec:framework}). 

\vspace{0.5mm} \noindent
(2) {\bf Effective RAG techniques.} 
We introduce a document retrieval mechanism addressing heterogeneous documents (Section~\ref{sec:docs_retrieval}) and design an effective method for telemetry analysis (Section~\ref{sec:telemetries_retrieval}). 

\vspace{0.5mm} \noindent
(3) 
{\bf Extensive experiments.} Experiments using real-world DBMS configuration debugging datasets show that \sys significantly outperforms existing alternatives. We make the code of \sys and the datasets in our experiments public on a Github repo\footnote{https://github.com/ruc-datalab/Andromeda/}.
\section{Preliminaries} \label{sec:pre}


\subsection{Problem Formalization} \label{subsec:problem}

\stitle{NL Debugging Question.}
Let $q$ be a user posed NL debugging question 
that reports unexpected configuration issues, which require diagnostic procedures over a DBMS ${D}$. Note that the question $q$ may express a variety of configuration debugging purposes, including but not limited to:
%
	\emph{(1) Diagnosing performance issues}, \eg much longer execution time that a database query or operation typically takes, and significant resource usage that the DBMS is using all available resources such as CPU, memory, or disk I/O. 
%
	\emph{(2) Fixing runtime errors}, \eg ``\emph{Error: Could not refresh instance}'' in MySQL. 
%
	\emph{(3) Customizing user-specific configurations}, \eg changing default decimal number and ignoring DBMS runtime warnings. 

Moreover, a question $q$ may also include the description of user's database (\eg schema and statistics), query workload and environments. For example, Figure~\ref{fig:overview} provides three NL debugging questions, which are, respectively, of the types of performance issues, runtime errors, and user-specific configurations.

\stitle{Configurable Knobs.}
Although there exist different ways to configure the DBMS~\cite{DBLP:conf/cidr/DiasRSVW05,DBLP:conf/sigmod/YoonNM16}, this paper focuses on diagnosing and tuning the DBMS \emph{configurable knobs} (or \emph{knobs} for short).
Specifically, modern DBMSs usually have hundreds of configurable knobs that control their runtime behavior, \eg $1,012$ knobs (such as \texttt{wait\_timeout}, \texttt{unique\_checks}) in MySQL 8.0 and more than $300$ knobs (such as \texttt{trace\_locks}) in PostgreSQL 12.0, which play crucial roles in DBMS tuning~\cite{tuneful,ottertune}.

Formally, we denote the set of configurable knobs in a DBMS as $\Kset$, where each $k \in \Kset$ is a knob with a possible value domain $\Dset_k$. Take the knob \texttt{autocommit} with value domain $\{0,1\}$ in MySQL as an example: If the \texttt{autocommit} mode is enabled (\ie \texttt{autocommit}=1), each SQL statement forms a single transaction on its own. If \texttt{autocommit} mode is disabled within a session (\ie \texttt{autocommit} = 0), the session always has a transaction open.


\stitle{DBMS Configuration Debugging.}
Given an NL debugging question $q$ over a DBMS $D$, the problem of DBMS configuration debugging aims to (1) diagnose a subset $K \subseteq \Kset$ of knobs relevant to the corresponding configuration issues in $q$, and (2) recommend a proper value $v \in \Dset_k$ for tuning each relevant knob $k \in K$. 
%
%
%
\begin{example}
	Figure~\ref{fig:running-example} illustrates a running example of DBMS configuration debugging. A database user poses an NL question $q$ to report a performance issue about a much longer execution time for an \texttt{INSERT} SQL statement (the left part of the figure). Given question $q$, the problem aims to diagnose a subset of troublesome configurable knobs, such as \texttt{foreign\_key\_checks} and \texttt{unique\_check} (the right part of the figure). Moreover, it also generates troubleshooting recommendations to fix the issues, \eg $\texttt{foreign\_key\_checks} = 0$ (\ie disabling foreign key check) and $\texttt{innodb\_buffer\_pool\_size}=2G$ (\ie increasing buffer pool size). 
%
\end{example}
\subsection{Related Work} \label{subsec:rw}

\stitle{DBMS Diagnosis and Configuration Tuning.}
Traditional DBMS diagnosis studies aim to identify and resolve performance and stability issues to ensure efficient, stable, and reliable operation of the underlying DBMS. These studies 
mainly employ techniques based on empirical rules, graph algorithms, and machine learning models~\cite{ADDM, DBSherlock, ISQUAD, DBLP:journals/corr/abs-2310-07637}. 

Recently, modern DBMSs are equipped with automatic configuration tuning tools~\cite{kpexp, ottertune, tuneful, iTuned, ResTune,DBLP:conf/sigmod/ZhangLZLXCXWCLR19, QTune, Feurer2018ScalableMF}. Some representative studies include the Lasso algorithm in OtterTune~\cite{ottertune}, Sensitivity Analysis (SA) in Tuneful~\cite{tuneful} for automatic knob selection, Bayesian optimization (\eg iTuned~\cite{iTuned} and ResTune~\cite{ResTune}), reinforcement learning (\eg CDBTune~\cite{DBLP:conf/sigmod/ZhangLZLXCXWCLR19} and Qtune~\cite{QTune}) for configuration optimization, and RGPE~\cite{Feurer2018ScalableMF} in ResTune~\cite{ResTune} for knowledge transformation. Recently, GPTuner~\cite{GPTuner} employs LLMs for configuration tuning based on DBMS manuals and Bayesian optimization.


However, these studies are not designed for taking NL questions from database users as input and recommend configurations corresponding to the issues, which is a key feature of our \sys framework. Moreover,  they are constructed by relying on pre-defined optimization objectives (such as runtime), making them difficult to address runtime errors and user-specific configurations.


\stitle{LLM-based DBMS Diagnosis.}
Panda~\cite{panda} takes the user's NL question as input, and introduces a context-grounding mechanism to LLMs by retrieving relevant documents and telemetries, so as to generate troubleshooting recommendations for the questions. Note that there are two differences between Panda and \sys. First, \sys explicitly outputs specific and accurate DBMS configuration settings with knobs and their values, whereas Panda outputs text-based debugging recommendations. Second, compared with Panda, \sys develops effective RAG techniques. 


D-Bot~\cite{dbot} is also a recent LLM-based DBMS diagnosis tool that acquires knowledge from diagnosis documents, and generates a diagnosis report (\textit{i.e.}, identifying the root causes and solutions) when a DBMS triggers an alert. There are two differences between D-Bot and \sys. The first is that the input of \sys is NL questions while that of D-Bot is triggered by alerts based on rules during DBMS execution. The second is that \sys outputs specific DBMS configuration settings with knob types and values, whereas D-Bot outputs analysis report texts that may not explicitly recommend specific knob configurations.



\stitle{Retrieval-Augmented Generation (RAG).} 
RAG has been proved to excel in many tasks, including open-ended question answering~\cite{siriwardhana2023improving, jeong2024adaptive}, programming context~\cite{autoformula, haipipe}, data matching\cite{unicorn} and fact verification~\cite{zamani2024stochastic, asai2023self, WebGPT, Meta}. 
Automatic DBMS configuration debugging is knowledge-extensive and requires experts to guarantee the accurate diagnosis of LLMs. In this paper, we introduce a framework called \sys based on RAG for configuration debugging, and demonstrate its effectiveness.

\revised{\stitle{Information Retrieval.} Several information retrieval (IR) techniques focus on systems that respond to questions by utilizing a community or a dataset of previously answered questions. Conventional methods employ neural networks \cite{cqann1}, knowledge graphs \cite{cqakg1, cqakg2}, and user intent analysis \cite{cqau1, cqau2, cqau3}. With the development of LLMs, more approaches consider incorporating LLMs to tackle the question answering problems~\cite{WebGPT, Meta}. \cite{Meta} introduces Question-Answer Cross Attention Networks (QAN) to leverage pre-trained models for answer selection, and improves the performance using knowledge augmentation from LLMs. Compared to these methods, the unique challenge of \sys lies in retrieving documents from multiple sources, such as historical questions and DBMS manuals, where the semantics of documents from different sources may be heterogeneous. For instance, 
	historical questions typically describe ``what is wrong''
	, while DBMS manuals focus on ``how to fix it''.
	Moreover, it is hard for existing IR techniques to analyze telemetry data in the DBMS debugging scenarios.}

\section{The \sys Framework}\label{sec:framework}

\begin{figure}[t]
	\centering
	\includegraphics[width=0.75\columnwidth]{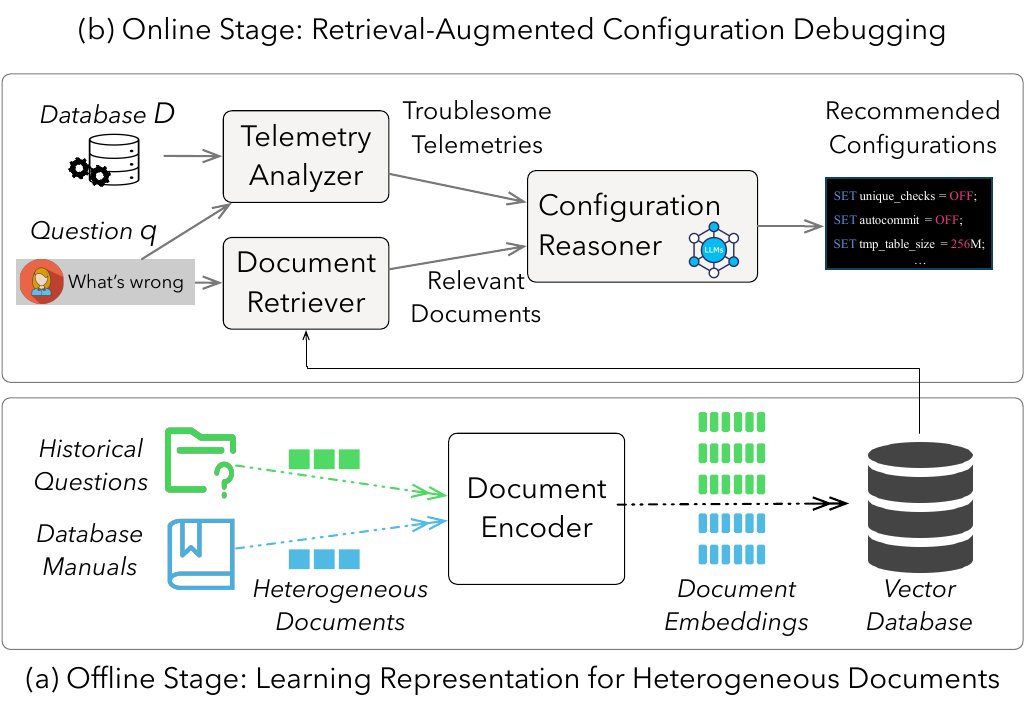}
	\caption{An overview of \sys. 
		(a) Offline: \sys learns representation for heterogeneous documents and stores the document embeddings in a vector database. 
		(b) Online: \sys utilizes an RAG-based configuration debugging strategy to recommend configurations for an NL question $q$ and a database $D$.
}
	%
\label{fig:framework}
\end{figure}

As shown in Figure~\ref{fig:framework},
\sys employs a retrieval-augmented generation (RAG) strategy consisting of offline and online stages. 

\stitle{Offline Stage}. In the offline stage, \sys learns a unified representation \ie \emph{a vector-based embedding} for heterogeneous documents from multiple sources. \sys currently considers the following two types of document sources. 
\begin{itemize}
	\item \emph{Historical Questions $\Qset$} contains historical configuration debugging questions that users have encountered, where each instance $\langle \qhis_i, K_i \rangle$ contains a historical question $\qhis_i$ and the recommended configurations $K_i$ corresponding to  $\qhis_i$. Figure~\ref{fig:running-example} shows a historical question regarding ``\emph{slow execution while inserting large amounts of data}'' with its recommended configurations, $\texttt{foreign\_key\_checks}=0$ (disabling foreign key checks) and $\texttt{unique\_checks}=0$ (disabling unique checks). 
	%
	\item \emph{Troubleshooting Manuals $\Mset$} consists of documents for DBMS manuals, \eg MySQL Manual~\cite{mso_manual} and PostgreSQL Manual~\cite{pg_manual}, which describe how to troubleshoot configuration issues. Specifically, we first divide each document in the manuals into sentences $\{\mhis_i\}$, and then use regular expression rules to match each sentence $\mhis_i$ with the recommended configurations $K_i$. Consider the example in Figure~\ref{fig:running-example} again. A document from \textit{MySQL Manual 10.5.5: Bulk Data Loading for InnoDB Tables} contains a sentence ``\emph{When importing data into InnoDB, turn off \texttt{autocommit} mode \ldots}'' with its corresponding configuration $\texttt{autocommit}=0$. In this way, we form a set of troubleshooting manuals $\Mset=\{\langle \mhis_i, K_i\rangle\}$.		
\end{itemize}

Given document sets $\Qset$ and $\Mset$, \sys uses a \textsc{Document Encoder} to map each document (either the historical question $\qhis$ or manual instance $\mhis$) into a vector-based embedding representation. Then, it stores the learned document embeddings in a vector database, which will be further used for online document retrieval. 

The key challenge here is that the semantics of documents from different sources may be heterogeneous. Consider the examples of historical questions and troubleshooting manuals in Figure~\ref{fig:running-example} again: historical questions typically describe ``\emph{what is wrong}'', while manuals focus on ``\emph{How to fix it}''. Thus, it is challenging to develop a unified document encoder that understands the semantics of different types of documents for effective document retrieval. 
To address this challenge, we propose an effective document encoding mechanism that addresses semantic heterogeneity among different sources. We propose a contrastive learning-based approach that aligns documents from different sources into a unified embedding space, and enables accurate document retrieval on the space. We also design a data synthesis strategy to address the difficulty of limited training data. 
We will provide more details in Section~\ref{sec:docs_retrieval}.

%

\stitle{Online Stage.} In the online stage, \sys utilizes a RAG strategy to recommend configurations $K$ for an NL question $q$ over a DBMS $D$, which consists of the following three steps. 

\etitle{Step 1: Document Retrieval.} \sys encodes the question $q$ into its embedding representation and employs a \textsc{Document Retriever} to retrieve top-$k$ historical questions or troubleshooting manuals that are most similar to $q$ in the embedding space. 


\etitle{Step 2: Telemetry Analysis.}
Inspired by the way a DBA performs configuration debugging, \sys also needs to know what is happening in the DBMS $D$. To this end, we propose to extract key features from the telemetry data, \ie time-series data of performance metrics (\eg CPU utilization). Formally, we define the telemetry data as a set $\Sset = \{\shis_i, \{t^{i}_j\}\}$ of time-series performance metrics, where each instance consists of a performance metric $\shis_i$ and a collection of timestamps $\{t^{i}_j\}$. Note that not all performance metrics or all timestamps of a particular metric may be relevant to the user's question $q$. Thus, \sys utilizes a \textsc{Telemetry Analyzer} to detect ``troublesome'' telemetries $\{\shis_i\}$ with the particular timestamps when a performance issue occurs (\eg the duration of full CPU usages). Moreover, \textsc{Telemetry Analyzer} also suggests the related knobs that need adjustment. Consider the example in Figure~\ref{fig:running-example}. 
\sys considers telemetries of DBMS $D$ and detects three troublesome performance metrics, \eg \texttt{innodb\_log\_write\_requests}, and suggests the related knobs that need adjustment, \eg \texttt{innodb\_log\_file\_size}.

\etitle{Step 3: Configuration Reasoning. }
With the above useful and contextual information, \sys employs a \textsc{Configuration Reasoner} that leverages an LLM (\eg GPT-4 or Llama 3) to diagnose improper DBMS configurations and generate accurate troubleshooting recommendations. Specifically, it generates a prompt that consists of three components: (1) task instruction, (2) retrieved data, including historical questions, troubleshooting manuals and relevant telemetries, and (3) user's NL question $q$.
Figure~\ref{fig:running-example} provides an example of the prompting mechanism (the central part of the figure).
After that, \sys prompts an LLM to generate configuration recommendations (as shown in the right part of Figure~\ref{fig:running-example}). In particular, we utilize a two-phase prompting strategy that respectively 
(1) identify a subset $K \subseteq \Kset$ of knobs relevant to the corresponding configuration issues in $q$, and (2) recommend a proper value $v \in \Dset_k$ for tuning each relevant knob $k \in K$, \revised{which makes sense because (i) Using RAG, LLM can obtain knowledge about the knobs. For example, LLM can set \texttt{autocommit=0} with manual, and it can set \texttt{innodb\_log\_file\_size = 256} if the similar question sets the same value. (ii) LLMs may learn some database related knowledge in pre-training or fine-tuning process, which is helpful for knob value decisions}. \revised{Note that there could be multiple ways, such as different tuning knobs, to fix the same issue. \sys uses the LLM to reason about the various tuning knobs and select the most appropriate set. Specifically, \sys generates a prompt that includes all the retrieved data and then calls the LLM to recommend the configurations. } 

\revised{To use \sys flexibly, users can selectively utilize the sources (for instance, if telemetry is unavailable, they can only consider documents). Moreover, users can use only Step 1 to identify the correct subset of knobs or perform an end-to-end process to predict the knob values.}

The key challenge on the online stage is how to detect troublesome telemetries. Specifically, there are many types of telemetries in a DBMS, and only a small number of telemetries are relevant to the question. Moreover, even given a relevant telemetry, only a small proportion of timestamps are useful. Thus, it is non-trivial to detect such troublesome telemetries that are relevant to a user's NL question. 
To address the challenge, we introduce a telemetry analysis method to identify troublesome telemetries that are relevant to a user's NL question. The method first utilizes a seasonal-trend decomposition mechanism to find troublesome telemetries, and then judiciously selects the telemetries relevant to a user's question. 
Please refer to Section~\ref{sec:telemetries_retrieval} for more details.

\revised{\stitle{Discussion on Generalizability.} \sys can be easily generalized to other DBMSes, as the algorithms for training data generation, model training, and evaluation are adaptable to new DBMSes. When applying \sys to a new DBMS, all that is required is the specification of the knobs and the DBMS manual. \sys can then use these inputs to automatically generate the augmented data and train a new model during the offline stage, while supporting the evaluation of the DBMS in the online stage.}
\section{Multi-Source Document Retrieval} \label{sec:docs_retrieval}

\begin{figure}[t]
	\subfigure[A straightforward learning method.]{
		\includegraphics[width=\columnwidth]{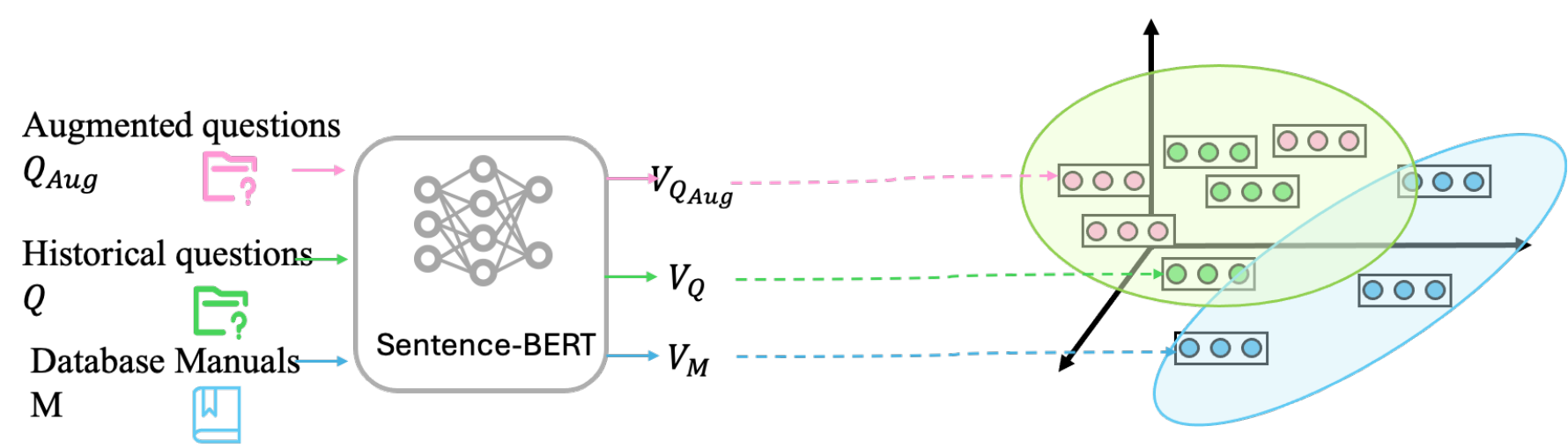}
		\label{fig:embedding-simple}
	}
	\subfigure[Our proposed contrastive learning approach.]{
		\includegraphics[width=\columnwidth]{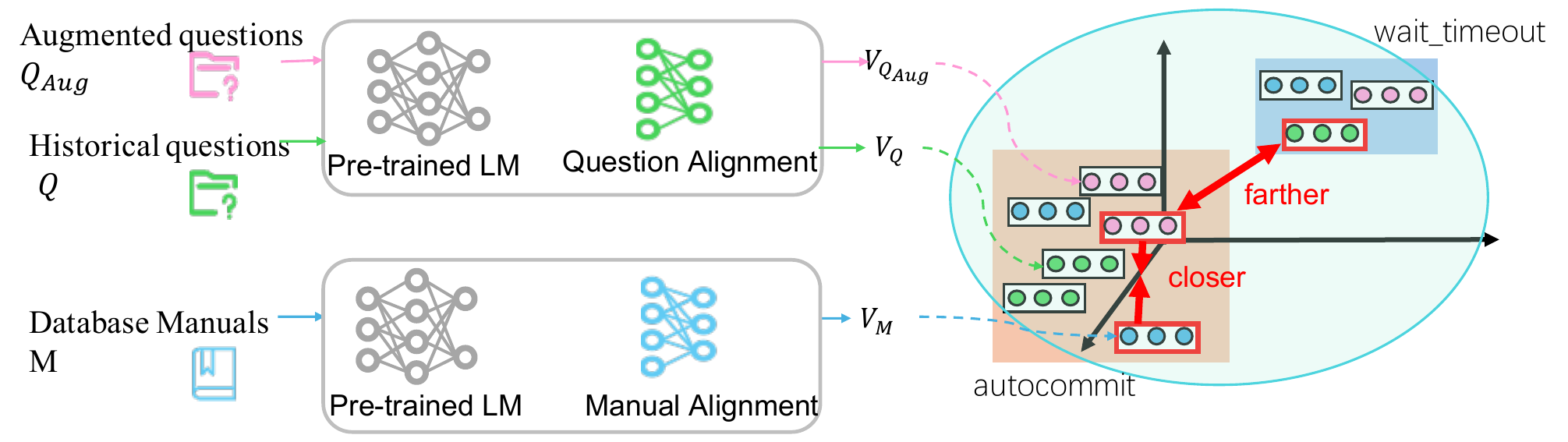}
		\label{fig:embedding-ours}		
	}
	\caption{An overview of document representation learning.}
	\label{fig:contrastive_learning}
\end{figure}

Given a user's NL question $q$, multi-source document retrieval aims to retrieve relevant historical questions from $\Qset$ and related troubleshooting manuals from $\Mset$.
To achieve this goal, we introduce an encoder model $\enc$ that maps the question $q$, historical questions $\Qset$ and DBMS manuals $\Mset$ into dense vectors, as shown in Figure~\ref{fig:contrastive_learning}.
Then, we can efficiently retrieve top-$k$ relevant documents by using the normalized L2 distance of the question $q$ and a document (\ie a historical question $\qhis$ or a manual instance $\mhis$). 
%

%
%
Intuitively, we develop techniques to address two main research challenges in designing encoder $\enc$. 

Firstly, as mentioned previously, manuals (usually including explanations of knobs and solutions) and questions (typically including configuration issues and user behaviors) are semantically heterogeneous. Thus, a straightforward encoding solution (\eg using Sentence-BERT~\cite{DBLP:conf/emnlp/ReimersG19}) may result in distributional divergence among documents from different sources, as shown in Figure~\ref{fig:contrastive_learning} (a). To address this challenge, we develop a contrastive learning approach that constructs \emph{comparative relationships} for documents from different sources, so as to \emph{align} the heterogeneous documents in a unified representation space, as shown in Figure~\ref{fig:contrastive_learning} (b). 

Secondly, constructing such an encoder $\enc$ that aligns documents from different sources requires a substantial amount of training data. However, in practice, there is often an insufficiency in both quantity and quality of the training data. To address the challenge, we propose a logic-chain based task decomposition method that uses LLMs to perform logical reasoning with the knowledge from manuals, thereby augmenting high-quality training data.

\subsection{Document Representation Learning} \label{sec:constractive_learning}
We present our contrastive learning approach that constructs \emph{comparative relationships} for documents from different sources.



\stitle{Model Structure.} We construct $\enc$ as a deep neural network. As shown in Figure~\ref{fig:contrastive_learning}, we design representation models separately for different document types, \ie questions and manuals. 
Specifically, for each document type, we firstly employ a pre-trained language model (LM) (\eg Sentence-BERT) to encode a document into a vector-based representation. Then, we add an \emph{Alignment} component to map the original embeddings into a unified representation space. In particular, the Alignment component is implemented using a neural network with multiple fully-connected layers. 


\stitle{Model Training.}
We design a contrastive learning approach to train the above model, \ie optimizing parameters for both pre-trained LMs and our Alignment components. To this end, we need to prepare both \emph{positive} and \emph{negative} documents for each target question $q^{*}$ in the training data, which are described as follows.

\etitle{(1) Positive document} is defined as the historical question $\qhis$ or manual instance $\mhis$ related to $q^{*}$, \ie the recommended knobs of $\qhis$ or $\mhis$ match that of $q^{*}$. To efficiently prepare positive documents, we first use Sentence-BERT to identify the similar documents from $\Qset$ and $\Mset$ to $q^{*}$, and then filter out the documents with knobs not overlapping with the recommended knobs of $q^*$. 

\etitle{(2) Negative document} is defined as the historical question $\qhis$ or manual instance $\mhis$ unrelated to $q^{*}$, \ie the recommended knobs of $\qhis$ or $\mhis$ are not overlapping with that of $q^{*}$. Similar to positive documents, we also use Sentence-BERT to identify the similar documents from $\Qset$ and $\Mset$ to $q^{*}$, and then only consider the documents with knobs not intersecting with that of $q^*$.

In such a way, we can prepare three types of training data instances.
(1) \emph{question-manual}: given a target question, the positive documents are historical questions and negative documents are manuals.
(2) \emph{question-question}: given a target question, both positive and negative documents are historical questions.
(3) \emph{manual-manual}:  given a target question, both positive and negative documents are manuals.
For each type, we prepare training data with an equal number of positive and negative documents. Additionally, we prioritize training manuals-manuals to achieve overall alignment of the manuals, and then proceed for other types.


For model training, 
we freeze the pre-trained LM (\ie Sentence-BERT in our implementation) and optimize the alignment models through contrastive learning, with the InfoNCE loss function,
\begin{equation}
L_{InfoNCE} = -\log \frac{\exp(||V_{q^{*}}, V_{d^{+}}||^2/\tau)}{\sum_{d^{-} \in D^{-}} \exp(||V_{q^{*}}, V_{d^{-}})||^2/\tau},
\end{equation}
where $d^{+}$ is positive document, while $d^{-} \in D^{-}$ are negative documents. The basic idea of the above InfoNCE loss brings positive retrieval texts closer in the embedding space while pushing negative retrieval texts farther apart, as shown in Figure~\ref{fig:contrastive_learning} (b).

\subsection{Training Data Augmentation} \label{sec:da}
For better training of $\enc$, we need \emph{a sufficient amount of training data}, \ie a large number of target queries ($q^{*}$) with ground-truth knob configurations. However, it is time- and effort-consuming to collect a large amount of high-quality training data, which hinders the effectiveness of document representation learning. To solve this problem, we introduce a novel \textbf{data synthesis} mechanism that utilizes an LLM to \emph{automatically generate NL questions with corresponding knob configurations from DBMS manuals}.

\begin{figure}[t]
	\subfigure[An example of logic-chain.]{
		\includegraphics[width=0.75\columnwidth]{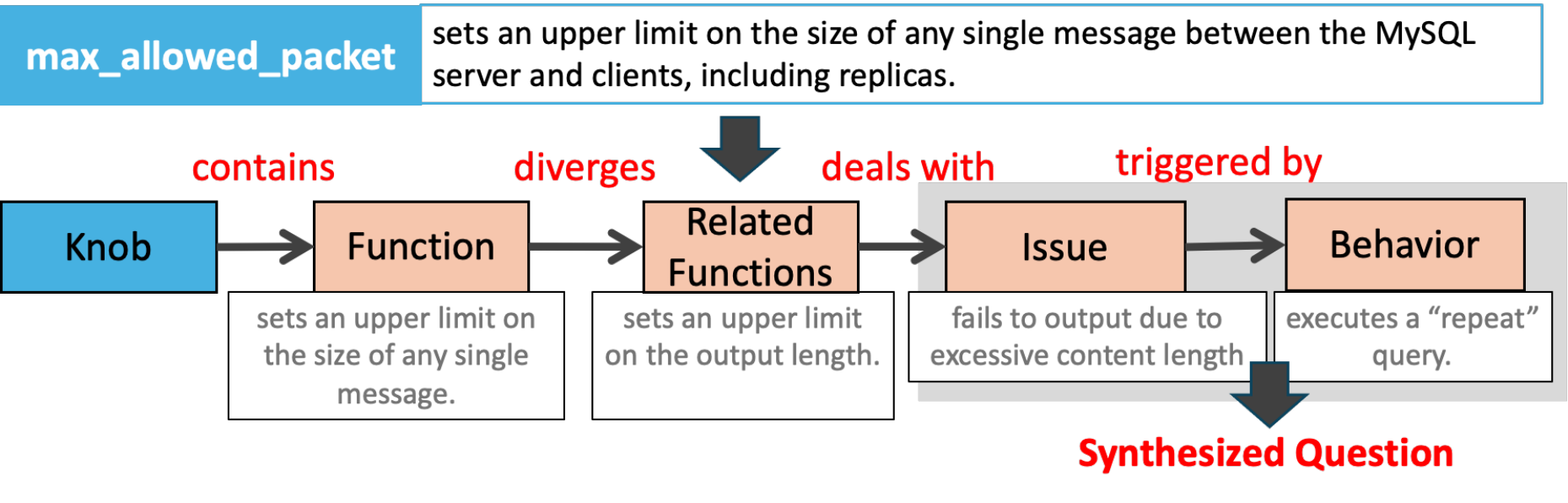}
		\label{fig:logicchian}
	}
	\subfigure[Data synthesis using a logic-chain strategy.]{
  \includegraphics[width=1\columnwidth]{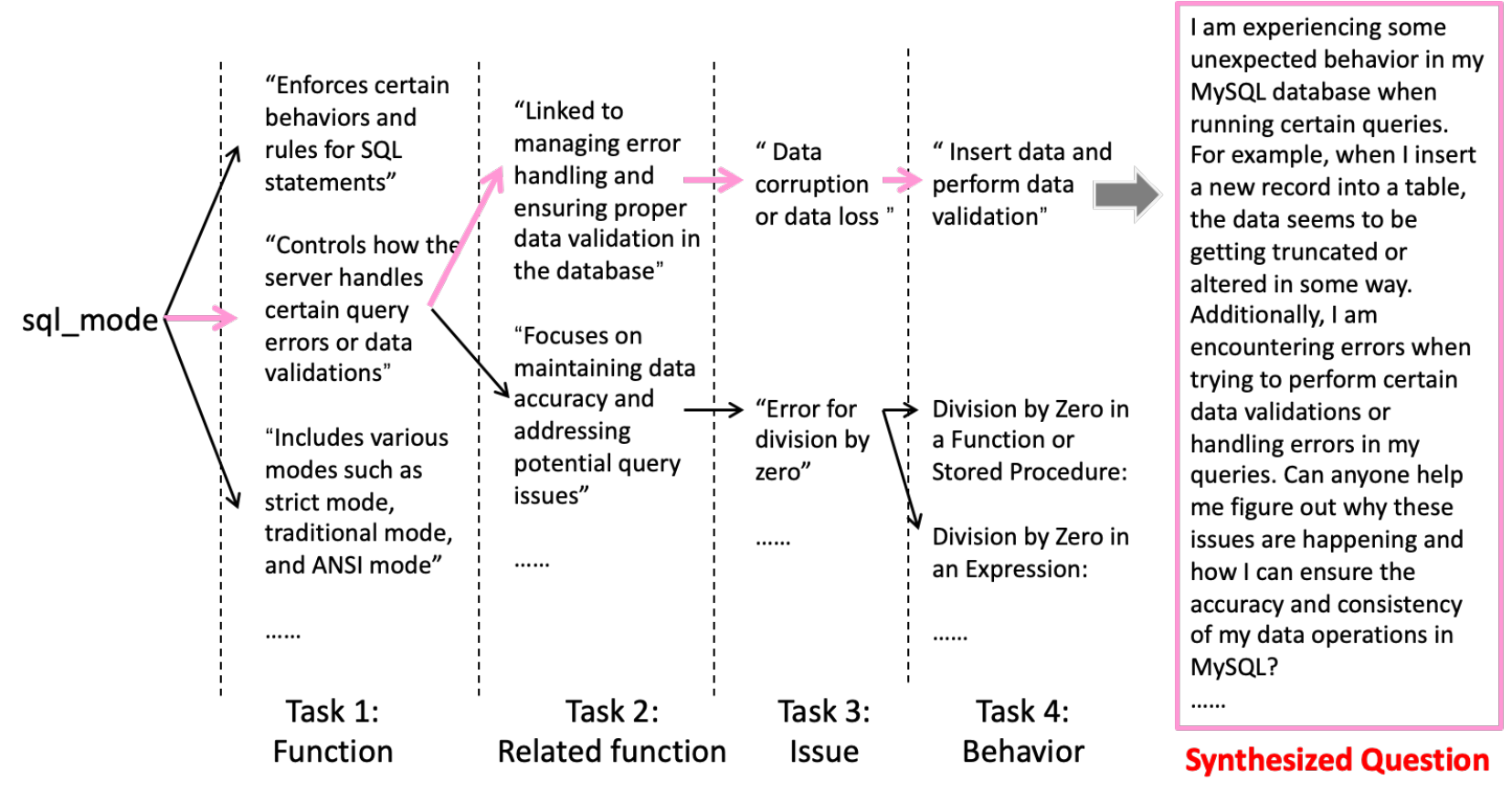}
		\label{fig:logicforest}
	}
	\caption{An overview of training data augmentation.}
\end{figure}

\begin{figure*}[t]
	\centering 
	\includegraphics[width=\textwidth]{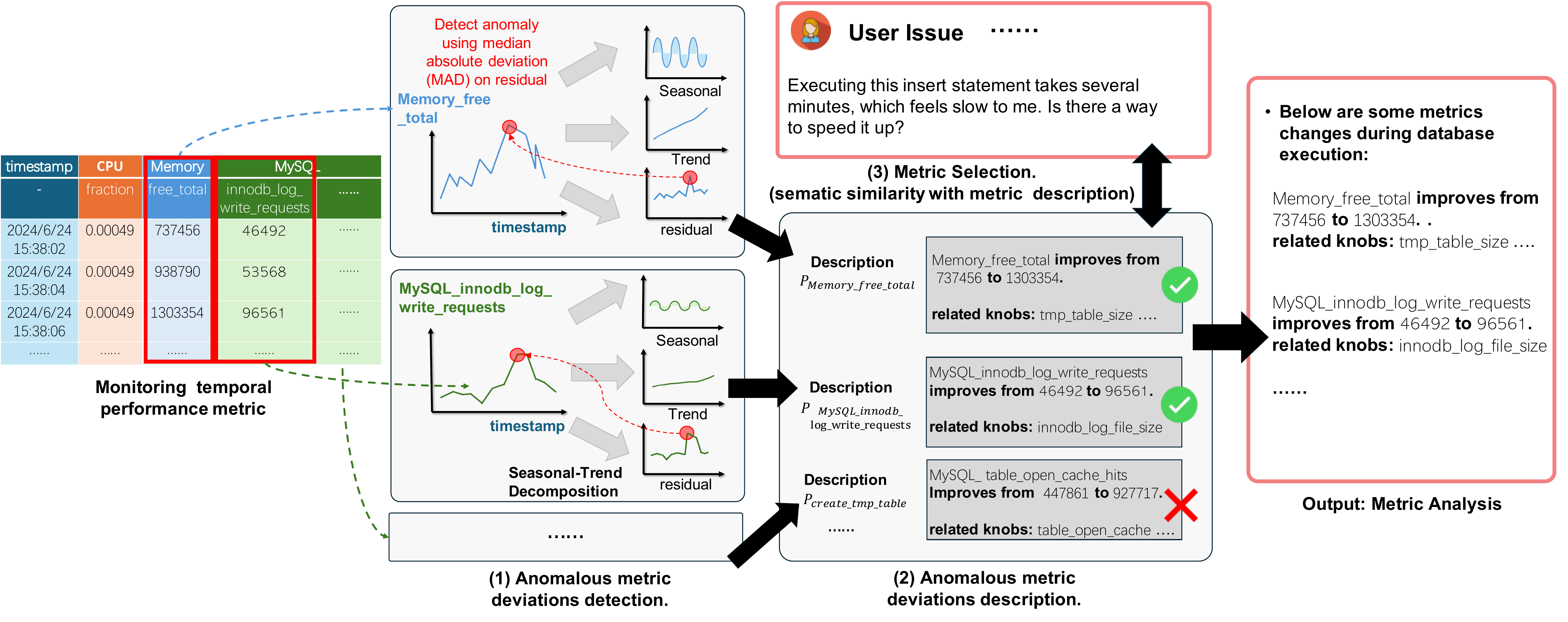}
	\caption{An overview of telemetry data analysis, which employs anomalous telemetry detection, telemetry-to-text, and relevant telemetry selection to report the ``anomalous'' status of database $D$.}
	\label{fig:metric_analysis} 
\end{figure*}


Unfortunately, a straightforward data synthesis method, which directly prompts the LLM with manuals and asks it to generate possible user's questions based on the manuals, may not perform well. The main reason is that the generated questions are not \emph{realistic}, \eg lacking contextual information, such as the user's behavior details. Moreover, the generated questions may also fall short of diversity. We illustrate this in the following example. 
\begin{example}[Straightforward Data Synthesis]
	Consider the manual of knob \texttt{ max\_allowed\_packet} shown in Figure~\ref{fig:logicchian}. Directly prompting an LLM (\eg GPT-4) returns the following question: ``\textit{I'm experiencing issues with my MySQL replication setup. Occasionally, large BLOB or TEXT fields cause errors on the source server, which then leads to the replication process stopping unexpectedly. What could be causing this issue and how can I resolve it?}''. Compared with the real questions shown in Figure~\ref{fig:overview}, it is not difficult to find such questions are not realistic.
\end{example}

%

%

\stitle{Our Logic-Chain Strategy.} 
To address the above limitation, we introduce a logic-chain based strategy for effective data synthesis. The basic idea is to generate realistic questions by simulating a real user's experience when encountering a DBMS configuration issue. Specifically, we consider the natural logical reasoning chain from issues to the solutions (\eg recommended knob configurations):

\texttt{Triggering {behaviors} $\rightarrow$ Behaviors causing {issues} $\rightarrow$ Specific {functions} resolving issues $\rightarrow$ {knobs} with the corresponding functions}. 

Intuitively, if we traverse the logic-chain reversely, we can inject domain knowledge about the logical relationships between knobs and issues into the LLM. Moreover, as the functions can be quite diverse, given a knob, we can generate a wide range of possible questions, which is helpful to improve the diversity of training data.


\begin{example}[Our Logic-Chain based Synthesis Strategy.]
Consider the manual of \texttt{max\_allowed\_packet} in the previous example again. 
We can extract diverse knob functions from the manual, and then we can infer that function ``\textit{set an upper limit on the output length}'' can resolve issue ``\textit{fail to output due to excessive content length}'', which is triggered by the behavior ``executes a repeat command''. We provide such a logic-chain to the LLM, and obtain a more realistic question ``\textit{I want to connect to the remote SQL server to simulate traffic using the REPEAT command in MySQL. And I executed SQL: REPEAT('A', 26214400). The connection is successful, but I encounter an error that returns NULL. How can I solve this for MySQL?}''
\end{example}

Based on the above idea, we utilize an LLM with a logic-chain prompting to generate questions. Specifically, we decompose the data synthesis task into several sub-tasks. Each sub-task aims to generate details for each edge in the logic-chain using the LLM, as shown in Figure~\ref{fig:logicforest}.
For instance, for ``\texttt{{knob}->{function}}'' in the logic-chain, we can construct the sub-task with question ``\textit{What are the functions of \texttt{max\_allowed\_packet}?}'', which can be answered as \textit{spec\_func} by the LLM. Next, for  ``\texttt{{function}->{Anomaly}}'', the question can be ``\textit{What anomaly can be solved with the {spec\_func} of \texttt{max\_allowed\_packet}}''. When the LLM provides answers for each edge in the complete logic-chain, we can integrate all the answers to identify a specific configuration debugging question.
\section{Telemetry Data Analysis} \label{sec:telemetries_retrieval}

%
We introduce a telemetry data analysis approach consisting of anomalous telemetry detection, telemetry-to-text and relevant telemetry selection. 
%
%
%
We develop techniques to address the following two challenges in the above three steps. 
The first challenge is detecting anomalous telemetries from a huge amount of time-series data, as it is not easy to identify what is anomalous. To address this, we first use a seasonal-trend decomposition method to decompose time series data and detect anomalous telemetries using statistics on the residual. Secondly, there may exist hundreds of telemetries, many of which may not be related to our NL question $q$. To tackle this challenge, we propose a telemetry-to-text mechanism that enables us to select telemetries relevant to $q$. 

%
%

\subsection{Anomalous Telemetry Detection} 

Given a time series $T_{s_i}$ of the telemetry metric $s_{i}$, the goal of this step is to detect a set of $\{\langle j, t_{j} \rangle \}$ where the value $t_{j}$ on timestamp $j$ in $T_{s_i}$ is anomalous (higher or lower than normal values). 
%
Inspired by~\cite{STL, oneshotstl}, for $T_{s_i}$, we use a seasonal-trend decomposition mechanism based on LOESS (STL)~\cite{STL}, which is a method that decomposes time series into seasonal, trend, and residual components for time series analysis, \ie 
\begin{equation}
T_{s_i} = Seasonal_{s_i} + Trend_{s_i} + \epsilon_{s_i}
\end{equation}
where $Seasonal_{s_i}$ represents the captured repeating seasonal patterns by smoothing the data within each period. $Trend_{s_i}$ is the long-term trend captured in the time series using local regression smoothing (LOESS), and $\epsilon$ is the unexplained random part of the time series after removing the trend and seasonal components. With STL, we can decompose seasonal and trends (normal variations), and assess whether the value on each timestamp is anomalous by analyzing the residual data $\epsilon_{si}$. After removing the effects of seasonal and trend components, it is easier to detect anomalous deviations for $s_i$ using statistic-based methods.

We use a statistic method called Extreme Studentized Deviate (ESD)~\cite{ESD} to detect anomalous deviations on $\epsilon_{si}$. The advantage of ESD is that it does not require pre-setting fixed thresholds or critical values for anomaly detection. Instead, it dynamically determines anomalous thresholds based on the actual distribution of data, which facilitates its adaptability across different temporal metrics in our scenarios. Specifically, ESD is an iterative algorithm to detect anomalous deviations. In each iteration $k$, we need to select an anomalous timestamp by calculating a deviation score $C_{t_j}$ for the value $t_j$ in each timestamp $j$ and a threshold $\lambda_k$. We select the timestamp $j_{max}$ with the highest deviation score $C_{max}$ and compare it with the threshold $\lambda_k$. If $C_{max} > \lambda_k$, $t_{j_{max}}$ is anomalous and we drop it.  We define the deviation score $C_{t_j}$ as:
%
%
\begin{equation}
C_{t_j} = \frac{|\epsilon_{t_{j}} - \tilde{\epsilon_{s_i}}|}{MAD_{j}}
\label{equ:deviation}
\end{equation}
%
Different from using mean and standard deviation, we use median and the median absolute deviation (MAD)~\cite{MAD} to calculate $C_{t_j}$, as they are robust to outliers~\cite{robust_stat}. In particular, $\tilde{\epsilon_{s_i}}$ in Equation~(\ref{equ:deviation}) is the median of $\epsilon_{s_i}$, the residual $\epsilon_{t_{j}}$ for timestamp $j$ in $\epsilon_{s_i}$ is:
\begin{equation}
\epsilon_{t_{j}} = t_j - Seasonal_{{s_i}_j}  - Trend_{{s_i}_j}
\end{equation}
The median absolute deviation (MAD) is the median of the absolute deviations from the median of all data points in $\lambda_{s_i}$, denoted by:

\begin{equation}
    MAD_{j} = median_j(|\epsilon_{j} - \tilde{\epsilon_{si}}|)
    \label{equ:MAD}
\end{equation}

After calculating the deviation $C_{t_j}$, we need to calculate the critical value $\lambda_k$ of $k^{th}$ iteration as threshold:
\begin{equation}
    \lambda_k = \frac{(n - k) t_{p, n-k-1}}{\sqrt{(n - k - 1 + t^2_{p, n-k-1})(n - k + 1)}} 
    \label{equ:lambda}
\end{equation}
where $t_{p, n-k-1}$ is the critical value of the t-distribution. If the the point with the highest deviation exceeds the critical value $\lambda_k$, that point is considered an anomaly and be dropped. The process is then repeated until no new anomalies are detected or the predetermined limit on the number of anomalies $N$ is reached and we can detect all the anomalous telemetries as $A$.

\subsection{Telemetry-to-Text and Telemetry Selection} 

After detecting anomalous telemetries $A$, for each metric $s_i$ in $T_{s_i}$, we would like to convert these anomalies into NL descriptions $p_{s_i}$. In particular, each description $p_{s_i} \in p_{s_i}$ should include the metric, an explanation of the metric, a description of the deviation (including the normal value before the deviation and the abnormal value after the deviation) and the corresponding related knobs. The reason for such a telemetry-to-text process is twofold. Firstly, it is helpful to identify the relevance between $p_{s_i}$ and the user's question $q$. Secondly, converting it into natural language enables the LLM to understand the information about users' DBMS status.

To this end, we first need to identify the normal values of each detected metric $s_i$ using a percentile point of the time series to represent the normal point (e.g. 5\%). In addition, we also attach the corresponding knobs that can adjust the metric $s_i$.
%
%
Using this Telemetry-to-Text process, we can obtain many telemetry descriptions in NL, such as "\textit{the value of an \texttt{innodb\_log\_write\_requests} changed from $46492$ to $96561$ and the related knob is \texttt{innodb\_log\_file\_size}.}"

 Due to the input-length limitations of LLM and the possibility that irrelevant metrics may mislead the LLM, we need to select useful anomalous telemetries from a large set as prompts to assist LLM to perform reasoning of DBMS diagnosis. Intuitively, we need to select metrics that are relevant to the user's question $q$. To achieve this goal, we rank the anomalous telemetry descriptions on the basis of the semantic similarity between the descriptions and the user's NL question (\eg using Sentence-BERT~\cite{sbert}). Finally, we select the top-$K$ anomalous descriptions as the final prompts. 




\begin{example}
    As shown in Figure~\ref{fig:metric_analysis}, given a user's NL question about the slow execution time of ``\texttt{INSERT}'' statement, we monitor the temporal performance $T_S$ as a temporal telemetry data with hundreds of metrics about CPU, Memory and MySQL DBMS. For time-series data $T_{s_i}$ of each metric $s_i$, such as \texttt{Memory\_free\_total} recording the remaining memory, we iteratively detect the most anomalous value in $T_{s_i}$. In each iteration, we use STL to decompose $T_{s_i}$ as seasonal, trend and residual. For residual, we use the mean absolute deviation (MAD) to assess the severity of the anomaly at each timestamp $j$ and collect them. After all anomalous deviation collected (\texttt{Memory\_free\_total}, \texttt{MySQL\_innodb\_log\_wirte\_requests} and others), we convert them into text and select relevant telemetries using semantic similarity with question $q$. Finally, we concatenate the generated text as the result of telemetry data analysis (the right part of the figure). 
\end{example}

\section{Experiments} \label{sec:exp}


    

    





\subsection{Experimental Setup} \label{sec:exp-setup}

\stitle{DBMS and Configurable Knobs.}
We conduct our experiments on MySQL 5.7 and PostgreSQL (PG) 12.0. MySQL contains $582$ knobs and PostgreSQL contains $202$ knobs.

\stitle{Datasets.} 
We use four real-world datasets in two types of experimental settings ({\bf NL} and {\bf Runnable}). Table~\ref{tab:dataset} provides statistics of the datasets. \revised{Note that, the training labels can be automatically annotated. For the original training data, we use regular expressions to extract relevant knobs from accepted answers or comments on questions from StackOverflow or the MySQL Forum. For the augmented training data, we generate questions based on a knob's manual, with the knob serving as the label corresponding to the questions. To validate labeling quality, we have randomly sampled some training and testing cases for manual verification, which have confirmed that our automatic labeling is accurate.}


\etitle{(1) The NL Setting}: 
We use DBMS configuration debugging questions posted by real users in StackOverflow~\cite{stackoverflow_dataset} and MySQL community~\cite{mysql_forum}. Specifically, we consider the following three datasets. 
\begin{itemize}
	\item \textbf{MySQL Stackoverflow (MySQL SO)} contains real questions regarding MySQL DBMS from StackOverflow~\cite{stackoverflow_dataset}.
	\item \textbf{MySQL Forum} contains questions about MySQL DBMS from MySQL community~\cite{mysql_forum}.
	\item \textbf{PostgreSQL Stackoverflow (PG SO)} contains questions about PostgreSQL DBMS from StackOverflow~\cite{stackoverflow_dataset}.
\end{itemize}

For each of the above datasets, each instance $\langle q, K_q \rangle$ consists of an NL question $q$ from real-world users and the ground-truth knob configuration $K_q$. Note that the ground-truth $K_q$ is manually extracted from the accepted answers or comments in Stackoverflow or MySQL community. 
In addition, for each dataset, we split all instances $\{\langle q, K_q \rangle\}$ with a ratio of 7:2:1  into three subsets, namely historical questions, training data and test data. We also merge historical questions of MySQL SO and MySQL Forum to form a more complete historical question set for these two datasets. 



\etitle{(2) The Runnable Setting}: 
One limitation of the above NL setting is that it takes the accepted answers as ground-truth configurations and cannot evaluate whether the configuration issues are solved. Moreover, the NL setting does not have the telemetry data of the underlying DBMS. 
Thus, we also consider a \emph{runnable} setting. In this setting, 
%
%
besides providing an NL question $q$, users should also give the telemetry data $\Sset$ of their DBMS. To this end, we derive a dataset \textbf{MySQL Run} from the above MySQL datasets. In particular, we prepare a test set in \textbf{MySQL Run} by selecting \revised{$70$} test questions that can reproduce the DBMS runtime environment from the MySQL dataset in the NL setting. Moreover, we manually design $30$ NL debugging questions with runtime environments.
%


\stitle{Evaluation Methods.} We conduct two evaluation methods for the NL and Runnable setting respectively.



\etitle{(1) Evaluation for the NL setting}. 
We evaluate the performance of knob diagnosis of various algorithms. For each test question $q$, given a set of knobs $K_{A}$ predicted by an algorithm $A$, we compare $K_{A}$ with ground-truth $K_{q}$ and calculate the Recall, Precision and F1-score of question $q$. 
F1-score is computed as $2\cdot P\cdot R/(P+R)$, where precision $P$ is the proportion of predicted correct knobs to all predicted matching pairs, recall $R$ is the proportion of predicted correct knobs to all correct knobs. Then, for the entire test dataset, we compute the \emph{average F1-score} to evaluate performance.
%
%
%

\etitle{(2) Evaluation for the Runnable setting}. We evaluate the performance of DBMS configuration debugging of various approaches. For each test instance with DBMS $D$, after an algorithm $A$ predicts a set of knobs $K_A$, the algorithm also predicts value $v$ for each knob $(k = v), k \in K_A$, which are used to tune DBMS $D$. 
%
Then, we employ DBMS experts to verify whether the recommended configurations successfully solve the issue. 
Next, we introduce \emph{SuccessRate}, which is the ratio of successful configurations that solve the issues over all the issues, $SuccessRate=  \frac{\sum_{q \in Q_{\tt Test}} \text{Solved}(q)}{|Q_{\tt Test}|}$.

%
%
%


\stitle{Methods Compared}. We compare the following methods.


\etitle{PLM}. For pre-trained language model (PLM), we use Sentence-BERT~\cite{sbert} with the published model~\cite{sbert_model} as the base model, which takes NL questions as input. Moreover, we add an output layer that outputs a vector of size $|K|$, where $K$ represents all knobs. Each dimension of this vector represents the score that the corresponding knob should be recommended. 
\revised{We train \textbf{PLM} using the training data and \textbf{PLM+DA} using augmented training data in the NL setting. During the inference phase, we select the set of knobs whose output values are higher than $0.5$.}



\revised{\etitle{LLMs}. For large language models, we consider open-source and closed-source LLMs. For open-source LLMs, we evaluate \textbf{Llama8B} (Meta-Llama-3-8B-instruct\cite{lm}), \textbf{Llama70B} (Meta-Llama-3-70B-instruct\cite{llamab}), \textbf{Qwen7B} (Qwen2-7B-Instruct\cite{qw}) and \textbf{Qwen72B} (Qwen2-72B-Instruct\cite{qwenb}). For closed-source LLMs, we evaluate \textbf{GPT-3.5} and \textbf{GPT-4}. For each LLM, prompt contains a task instruction and an NL question. Moreover, we ask the LLM to strictly output a knob list so that we can extract the knob set for evaluation. For value recommendation, we ask LLMs to strictly output a dictionary as ${{\tt knob}:{\tt value}}$. If the LLMs do not output the correct format, we regard it as failures.}

\revised{\etitle{LLMs (all knobs)}. LLMs (all knobs) improves the LLMs by modifying the prompts, \ie adding all knobs $\Kset$ to the original prompt and requiring the model to select among them. We also ask the LLMs to output a knob list
	and dictionary for value recommendation.}

\revised{\etitle{LLMs (prompt engineering)}. We use two prompting strategies: (1) \textbf{CoT}, which asks the LLM to think step by step, and (2) \textbf{Task Decomposition}, which asks the LLM to decompose the entire process into multiple steps and solve them in a divide-and-conquer manner.}

\revised{\etitle{LLMs (SFT)}. We use the training data to construct the ``Instruction-Answer'' pairs for fine-tuning \texttt{gpt-3.5-turbo}, where the ``Instruction'' is the prompts in the \textit{LLM} setting, and the ``Answer'' consists of a list of ground-truth knobs from the training data. For inference, we apply the same prompting strategies in the \textit{LLMs} setting.}

\stitle{Implementation}. All experiments were conducted on a device with Linux Ubuntu 20.04.1, 20 vCPU cores and 1.0TB memory. For document retrieval, we use top-$5$ retrieved documents (questions \& manuals). For GPTs, we use Python to call OpenAI's API, where the versions are gpt-3.5-turbo for GPT-3.5 and gpt-4-turbo GPT-4 with $temperature = 0$ and $random\_seed=42$. For telemetry data analysis, we use Prometheus to monitor
$557$ performance metrics with a timestamp interval of $1$ second. \revised{We use Meta Faiss as our vector database to index the embeddings of historical questions and manuals. In the current implementation of \sys, we have indexed 5,138 vectors for MySQL and 1,158 vectors for PostgreSQL, with each vector having a dimension of 768}. 


\begin{table}[t!]
\caption{\revised{Statistics of Datasets, where $Q$ means historical questions and $M$ means DBMS manuals for retrieval, ``\# test/train'' means the number of question in test/train set and ``\% test/$|Q|$'' means the coverage ratio of knobs in test/historical questions set. Moreover, $|M|$ covers all knobs.}}
\resizebox{0.8\columnwidth}{!}
{

\revised{
\begin{tabular}{|c||c|c|c|c|c|c||c|}
\hline
\textbf{Datasets} & \textbf{\# test} & \textbf{\# train} & \textbf{$|Q|$} & \textbf{$|M|$} & \textbf{\% test} & \textbf{\% $|Q|$} & \textbf{Type} \\ \hline \hline
MySQL SO          & 174                 & 620                  & 1632           & 3506           & 30\%                       & 25\%                     & NL            \\ \hline
MySQL Forum       & 49                  & 620                  & 1632           & 3506           & 17\%                       & 25\%                     & NL            \\ \hline
PG SO             & 57                  & 114                  & 402            & 756            & 38\%                       & 19\%                     & NL            \\ \hline
MySQL Run         & 100                  & 620                  & 1632           & 3506           & 15\%                       & 25\%                     & Runnable      \\ \hline
\end{tabular}
}
}
\label{tab:dataset}
\end{table}

\subsection{Overall Comparisons} \label{sec:exp-overall}


\begin{table*}
\caption{\revised{
Overall results, where we measure Precision, Recall and F1-score in the NL setting, and SuccessRate in the Runnable setting.
}}
\revised{
\resizebox{1.\linewidth}{!}{
\begin{tabular}{|c|c||ccc|ccc|ccc||c|}
\hline
\multirow{2}{*}{\textbf{Methods}}                                            & \multirow{2}{*}{\textbf{Settings}} & \multicolumn{3}{c|}{\textbf{MySQL SO}}                                                             & \multicolumn{3}{c|}{\textbf{MySQL Forum}}                                                          & \multicolumn{3}{c||}{\textbf{PG SO}}                                                                & \textbf{MySQL Run}   \\ \cline{3-12} 
                                                                             &                                    & \multicolumn{1}{c|}{\textbf{Recall}} & \multicolumn{1}{c|}{\textbf{Precision}} & \textbf{F1-score} & \multicolumn{1}{c|}{\textbf{Recall}} & \multicolumn{1}{c|}{\textbf{Precision}} & \textbf{F1-score} & \multicolumn{1}{c|}{\textbf{Recall}} & \multicolumn{1}{c|}{\textbf{Precision}} & \textbf{F1-score} & \textbf{SuccessRate} \\ \hline\hline
\multirow{2}{*}{PLM}                                                         & PLM                                & \multicolumn{1}{c|}{0.529}           & \multicolumn{1}{c|}{0.004}              & 0.008             & \multicolumn{1}{c|}{0.547}           & \multicolumn{1}{c|}{0.008}              & 0.015             & \multicolumn{1}{c|}{\textbf{0.558}}  & \multicolumn{1}{c|}{0.003}              & 0.006             & -                    \\ \cline{2-12} 
                                                                             & PLM+DA                             & \multicolumn{1}{c|}{0.518}           & \multicolumn{1}{c|}{0.004}              & 0.008             & \multicolumn{1}{c|}{\textbf{0.587}}  & \multicolumn{1}{c|}{0.009}              & 0.017             & \multicolumn{1}{c|}{0.485}           & \multicolumn{1}{c|}{0.005}              & 0.01              & -                    \\ \hline\hline
\multirow{6}{*}{LLMs}                                                        & GPT-3.5                            & \multicolumn{1}{c|}{0.329}           & \multicolumn{1}{c|}{0.205}              & 0.221             & \multicolumn{1}{c|}{0.207}           & \multicolumn{1}{c|}{0.145}              & 0.153             & \multicolumn{1}{c|}{0.32}            & \multicolumn{1}{c|}{0.183}              & 0.201             & 0.37                 \\ \cline{2-12} 
                                                                             & GPT-4                              & \multicolumn{1}{c|}{0.291}           & \multicolumn{1}{c|}{0.201}              & 0.222             & \multicolumn{1}{c|}{0.143}           & \multicolumn{1}{c|}{0.052}              & 0.06              & \multicolumn{1}{c|}{0.272}           & \multicolumn{1}{c|}{0.14}               & 0.161             & 0.43                 \\ \cline{2-12} 
                                                                             & Llama8B                            & \multicolumn{1}{c|}{0.07}            & \multicolumn{1}{c|}{0.025}              & 0.035             & \multicolumn{1}{c|}{0.078}           & \multicolumn{1}{c|}{0.021}              & 0.033             & \multicolumn{1}{c|}{0.056}           & \multicolumn{1}{c|}{0.023}              & 0.032             & 0.25                 \\ \cline{2-12} 
                                                                             & Llama70B                           & \multicolumn{1}{c|}{0.054}           & \multicolumn{1}{c|}{0.027}              & 0.034             & \multicolumn{1}{c|}{0.051}           & \multicolumn{1}{c|}{0.026}              & 0.034             & \multicolumn{1}{c|}{0.091}           & \multicolumn{1}{c|}{0.031}              & 0.045             & 0.26                 \\ \cline{2-12} 
                                                                             & Qwen7B                             & \multicolumn{1}{c|}{0.024}           & \multicolumn{1}{c|}{0.024}              & 0.019             & \multicolumn{1}{c|}{0.008}           & \multicolumn{1}{c|}{0.006}              & 0.007             & \multicolumn{1}{c|}{0.044}           & \multicolumn{1}{c|}{0.015}              & 0.021             & 0.11                 \\ \cline{2-12} 
                                                                             & Qwen72B                            & \multicolumn{1}{c|}{0.208}           & \multicolumn{1}{c|}{0.125}              & 0.140             & \multicolumn{1}{c|}{.0.171}          & \multicolumn{1}{c|}{0.112}              & 0.116             & \multicolumn{1}{c|}{0.114}           & \multicolumn{1}{c|}{0.05}               & 0.065             & 0.31                 \\ \hline\hline
\multirow{6}{*}{\begin{tabular}[c]{@{}c@{}}LLMs \\ (all knobs)\end{tabular}} & GPT-3.5                            & \multicolumn{1}{c|}{0.448}           & \multicolumn{1}{c|}{0.293}              & 0.308             & \multicolumn{1}{c|}{0.411}           & \multicolumn{1}{c|}{0.247}              & 0.246             & \multicolumn{1}{c|}{0.465}           & \multicolumn{1}{c|}{0.268}              & 0.299             & 0.52                 \\ \cline{2-12} 
                                                                             & GPT-4                              & \multicolumn{1}{c|}{0.507}           & \multicolumn{1}{c|}{0.300}              & 0.332             & \multicolumn{1}{c|}{0.371}           & \multicolumn{1}{c|}{0.227}              & 0.235             & \multicolumn{1}{c|}{0.504}           & \multicolumn{1}{c|}{0.38}               & 0.390             & 0.49                 \\ \cline{2-12} 
                                                                             & Llama8B                            & \multicolumn{1}{c|}{0.116}           & \multicolumn{1}{c|}{0.048}              & 0.057             & \multicolumn{1}{c|}{0.11}            & \multicolumn{1}{c|}{0.054}              & 0.058             & \multicolumn{1}{c|}{0.263}           & \multicolumn{1}{c|}{0.094}              & 0.121             & 0.31                 \\ \cline{2-12} 
                                                                             & Llama70B                           & \multicolumn{1}{c|}{0.536}           & \multicolumn{1}{c|}{0.217}              & 0.268             & \multicolumn{1}{c|}{0.368}           & \multicolumn{1}{c|}{0.166}              & 0.201             & \multicolumn{1}{c|}{0.488}           & \multicolumn{1}{c|}{0.25}               & 0.277             & 0.57                 \\ \cline{2-12} 
                                                                             & Qwen7B                             & \multicolumn{1}{c|}{0.103}           & \multicolumn{1}{c|}{0.086}              & 0.087             & \multicolumn{1}{c|}{0.136}           & \multicolumn{1}{c|}{0.164}              & 0.137             & \multicolumn{1}{c|}{0.232}           & \multicolumn{1}{c|}{0.247}              & 0.222             & 0.10                 \\ \cline{2-12} 
                                                                             & Qwen72B                            & \multicolumn{1}{c|}{0.451}           & \multicolumn{1}{c|}{0.3}                & 0.321             & \multicolumn{1}{c|}{0.384}           & \multicolumn{1}{c|}{0.248}              & 0.26              & \multicolumn{1}{c|}{0.521}           & \multicolumn{1}{c|}{0.293}              & 0.305             & 0.54                 \\ \hline\hline
\multirow{2}{*}{\begin{tabular}[c]{@{}c@{}}LLMs\\ (PE)\end{tabular}}         & CoT                                & \multicolumn{1}{c|}{0.308}           & \multicolumn{1}{c|}{0.170}              & 0.196             & \multicolumn{1}{c|}{0.161}           & \multicolumn{1}{c|}{0.115}              & 0.121             & \multicolumn{1}{c|}{0.226}           & \multicolumn{1}{c|}{0.144}              & 0.158             & 0.53                 \\ \cline{2-12} 
                                                                             & Task Dec.                          & \multicolumn{1}{c|}{0.143}           & \multicolumn{1}{c|}{0.136}              & 0.127             & \multicolumn{1}{c|}{0.111}           & \multicolumn{1}{c|}{0.160}              & 0.100             & \multicolumn{1}{c|}{0.207}           & \multicolumn{1}{c|}{0.169}              & 0.179             & 0.5                  \\ \hline\hline
LLM (SFT)                                                                    & GPT-3.5                            & \multicolumn{1}{c|}{0.345}           & \multicolumn{1}{c|}{0.391}              & 0.347             & \multicolumn{1}{c|}{0.333}           & \multicolumn{1}{c|}{0.6}                & 0.390             & \multicolumn{1}{c|}{0.343}           & \multicolumn{1}{c|}{0.383}              & 0.347             & 0.46                 \\ \hline\hline
\multirow{6}{*}{Andromeda}                                                   & GPT-3.5                            & \multicolumn{1}{c|}{0.512}           & \multicolumn{1}{c|}{0.318}              & 0.348             & \multicolumn{1}{c|}{0.541}           & \multicolumn{1}{c|}{0.383}              & 0.382             & \multicolumn{1}{c|}{0.461}           & \multicolumn{1}{c|}{0.305}              & 0.321             & \textbf{0.79}        \\ \cline{2-12} 
                                                                             & GPT-4                              & \multicolumn{1}{c|}{\textbf{0.557}}  & \multicolumn{1}{c|}{0.426}              & \textbf{0.441}    & \multicolumn{1}{c|}{0.528}           & \multicolumn{1}{c|}{\textbf{0.495}}     & \textbf{0.449}    & \multicolumn{1}{c|}{0.47}            & \multicolumn{1}{c|}{0.402}              & 0.398             & 0.76                 \\ \cline{2-12} 
                                                                             & Llama8B                            & \multicolumn{1}{c|}{0.32}            & \multicolumn{1}{c|}{0.183}              & 0.201             & \multicolumn{1}{c|}{0.441}           & \multicolumn{1}{c|}{0.244}              & 0.256             & \multicolumn{1}{c|}{0.39}            & \multicolumn{1}{c|}{0.241}              & 0.270             & 0.44                 \\ \cline{2-12} 
                                                                             & Llama70B                           & \multicolumn{1}{c|}{0.507}           & \multicolumn{1}{c|}{0.315}              & 0.345             & \multicolumn{1}{c|}{0.482}           & \multicolumn{1}{c|}{0.331}              & 0.347             & \multicolumn{1}{c|}{0.452}           & \multicolumn{1}{c|}{0.31}               & 0.335             & 0.60                 \\ \cline{2-12} 
                                                                             & Qwen7B                             & \multicolumn{1}{c|}{0.294}           & \multicolumn{1}{c|}{0.245}              & 0.242             & \multicolumn{1}{c|}{0.253}           & \multicolumn{1}{c|}{0.192}              & 0.195             & \multicolumn{1}{c|}{0.275}           & \multicolumn{1}{c|}{0.225}              & 0.222             & 0.46                 \\ \cline{2-12} 
                                                                             & Qwen72B                            & \multicolumn{1}{c|}{0.529}           & \multicolumn{1}{c|}{\textbf{0.435}}     & 0.437             & \multicolumn{1}{c|}{0.488}           & \multicolumn{1}{c|}{0.45}               & 0.412             & \multicolumn{1}{c|}{0.449}           & \multicolumn{1}{c|}{\textbf{0.415}}     & \textbf{0.410}    & 0.60                 \\ \hline
\end{tabular}

}}
\label{tab:overall}
\end{table*}

\stitle{Exp-1: How does \sys perform compared with LM-based methods?}
We report the overall results in Table~\ref{tab:overall}. 
%

\stitle{Evaluation in the NL Setting}.
\revised{We report the Recall, Precision and F1-scores of different approaches on the three datasets, MySQL SO, MySQL Forum and PG SO, in the NL setting, as shown in Table~\ref{tab:overall}.
}
%
%

The PLM-based method does not perform well due to the limited availability of training data.
\revised{Even though PLM+DA incorporates more training data, the improvement still remains marginal. This is because the PLM, with its fewer parameters, struggles to understand the context, particularly in longer and expert-domain texts.}

LLM-based methods (LLMs in Table~\ref{tab:overall}) perform better than PLM-based methods. 
However, they also have unsatisfactory performance. The main reason is hallucination: they often predict knobs that do not exist in reality, even for GPT-4. \revised{Moreover, GPT models outperform open-source LLMs, while smaller-sized models (Llama8B and Qwen7B) fail to deliver satisfactory performance.}

This phenomenon gets better with LLMs (all knobs), because this essentially turns a generation problem into a selection problem, which significantly reduces hallucinations. However, these methods still have limitations. In the database configuration debugging scenario, the database contains a large number of knobs with complex functions, and there are intricate logical relationships between the knobs and anomalies. To deal with such complex questions, LLM models still lack expert knowledge and experience. Additionally, we find that sometimes GPT-4 does not perform better than GPT-3.5 (\eg in the MySQL Forum), because GPT-4 may not strictly output results as our required format.


We can see that \sys outperforms other methods significantly. This is mainly because \sys employs a retrieval-augmented generation (RAG) strategy that 
enriches the NL debugging questions with domain-specific context drawn from multiple sources,
including historical questions and troubleshooting manuals. The results have shown that the RAG strategy can significantly improve the quality and relevance of its recommendations. \revised{Moreover, with \sys, the performance gap between open-source models and GPTs narrows. Larger-size open-source LLMs (Llama70B and Qwen72B) become comparable to GPTs due to the advantage of retrieved knowledge.}

\revised{The prompt engineering methods (CoT and Task Decomposition in Table~\ref{tab:overall}) are not very effective because the configuration debugging process doesn't always involve multiple steps for decomposition. Instead, it requires expert knowledge of DBMS.}
\revised{Additionally, the fine-tuning methods do not show significant improvement, primarily due to the limitations of the training data. Unlike other classification tasks, knob diagnosis requires predictions across hundreds of knobs with diverse functions, each needing substantial training data to achieve accurate results.}

\stitle{Evaluation in the Runnable setting.} \sys produces high-accuracy (over $0.7$) and outperforms other approaches, which means most cases are practically solved by \sys. Note that the result of PLM is not reported, because it predicts a very large number of knobs (more than $100$) and fails to solve the issues. 

In detailed analysis, the number of knobs predicted on average by \sys is $4.85$ with GPT-3.5 and $2.63$ with GPT-4, which means that users can solve most problems by adjusting just a few knobs. Interestingly, \sys achieves higher accuracy when using GPT-3.5 compared to GPT-4. This is partly attributed to GPT-4's tendency to deviate from our specified output requirements, and it produces fewer knobs.
In contrast, when the number of adjustable knobs is controlled, GPT-3.5 recommends more knobs, making it easier to find knobs that can solve the problem. Compared with other methods, \sys shows significant improvements compared with GPTs and GPTs (all knobs).

\stitle{Finding 1: \sys outperforms other LM-based methods in both NL and Runnable evaluation settings, demonstrating the effectiveness of our RAG strategy.}




\subsection{Evaluation on Document Retrieval} \label{sec:exp-doc}

In this section, we evaluate the impact of our document retrieval module. In addition to F1-score for end-to-end results, we also directly evaluate the retrieval quality. 
We measure the quality of the documents by comparing the knobs of the top-$k$ retrieved documents with the ground truth knobs, thereby eliminating the influence of the reasoner module on the evaluation. To achieve it, we utilize \textbf{Document retrieval recall} metric. For algorithm $A$ on test dataset $Q_{\tt Test}$, \textbf{Document retrieval recall} is calculated as:
$ Recall_{doc}(A, k, Q_{\tt Test}) = \frac{1}{|Q_{\tt Test}|} \sum_{q \in Q_{\tt Test}} \frac{|(\bigcup_{d \in D_{A, k, q}} K_d) \cap K_{q}|}{|K_q|} $,
\noindent where $D_{A, k, q}$ is the top-$k$ retrieved documents from retrieval data of $A$ on $Q$, and $K_d$ is the corresponding knobs of document $d$. 


\stitle{Exp-2: How does data augmentation work in document retrieval module?}
In order to see the effect of the augmentation module, we consider an \textbf{\sys-docs-only} version of our framework by disabling telemetry analysis. 
We compare \textbf{\sys-docs-only} with the following document augmentation methods: \textbf{(1) No Augment}: training model using the original training data (as shown in Table~\ref{tab:dataset}). 
\textbf{(2) CoT Augment}: let LLM generate the same volume questions from manuals using the CoT method.
%
The results are provided in Table~\ref{tab:augment_recall} for \textbf{Document retrieval recall} and Table~\ref{tab:augment_reasoning} for \textbf{F1-score}, which show that \sys-docs-only outperforms these two methods. \sys-docs-only and CoT Augment perform better than No Augment, showing that the new generated question-knob pairs are helpful. \sys-docs-only performs better than CoT Augment due to our proposed logic-chain strategy.

%
%
%

\stitle{Finding 2: The logic-chain based data synthesis strategy for training data augmentation is very helpful and outperforms other data augmentation methods.}

\begin{table*}
\caption{Document retrieval recall of different augmentation methods in document retrieval in the NL evaluation setting.}
\label{tab:augment_recall}
 \resizebox{\textwidth}{!}{
 \begin{tabular}{|c|cccc|cccc|cccc|}
\hline
\textbf{Datasets}             & \multicolumn{4}{c|}{\textbf{MySQL SO}}                                                                                          & \multicolumn{4}{c|}{\textbf{MySQL Forum}}                                                                                        & \multicolumn{4}{c|}{\textbf{PG SO}}                                                                                              \\ \hline
\textbf{top-$k$}                & \multicolumn{1}{c|}{\textbf{1}}     & \multicolumn{1}{c|}{\textbf{3}}     & \multicolumn{1}{c|}{\textbf{5}}    & \textbf{10}    & \multicolumn{1}{c|}{\textbf{1}}     & \multicolumn{1}{c|}{\textbf{3}}     & \multicolumn{1}{c|}{\textbf{5}}     & \textbf{10}    & \multicolumn{1}{c|}{\textbf{1}}     & \multicolumn{1}{c|}{\textbf{3}}     & \multicolumn{1}{c|}{\textbf{5}}     & \textbf{10}    \\ \hline \hline
No Augment                    & \multicolumn{1}{c|}{0.323}          & \multicolumn{1}{c|}{0.407}          & \multicolumn{1}{c|}{0.474}         & 0.55           & \multicolumn{1}{c|}{0.314}          & \multicolumn{1}{c|}{0.474}          & \multicolumn{1}{c|}{0.582}          & 0.63           & \multicolumn{1}{c|}{0.194}          & \multicolumn{1}{c|}{0.277}          & \multicolumn{1}{c|}{0.35}           & 0.407          \\ \hline
CoT Augment                   & \multicolumn{1}{c|}{0.32}           & \multicolumn{1}{c|}{0.435}          & \multicolumn{1}{c|}{0.516}         & 0.587          & \multicolumn{1}{c|}{0.305}          & \multicolumn{1}{c|}{0.431}          & \multicolumn{1}{c|}{0.524}          & 0.612          & \multicolumn{1}{c|}{0.247}          & \multicolumn{1}{c|}{0.373}          & \multicolumn{1}{c|}{0.454}          & \textbf{0.556} \\ \hline
\sys-docs-only & \multicolumn{1}{c|}{\textbf{0.327}} & \multicolumn{1}{c|}{\textbf{0.461}} & \multicolumn{1}{c|}{\textbf{0.54}} & \textbf{0.621} & \multicolumn{1}{c|}{\textbf{0.336}} & \multicolumn{1}{c|}{\textbf{0.492}} & \multicolumn{1}{c|}{\textbf{0.595}} & \textbf{0.672} & \multicolumn{1}{c|}{\textbf{0.299}} & \multicolumn{1}{c|}{\textbf{0.426}} & \multicolumn{1}{c|}{\textbf{0.473}} & 0.533          \\ \hline
\end{tabular}
}
\end{table*}

\begin{table}
\caption{Reasoning result on F1-score of augmentation methods, where No Aug. means No Augment, CoT Aug. means CoT Augment. and And-d means \sys-docs-only.}
\label{tab:augment_reasoning}
 \resizebox{0.55\columnwidth}{!}{
\begin{tabular}{|c|c||c|c|c|}
\hline
\textbf{Datasets}                                                       & \textbf{Reasoners} & \textbf{No Aug.} & \textbf{CoT Aug.} & \textbf{And-d} \\ \hline \hline
\multirow{2}{*}{\begin{tabular}[c]{@{}c@{}}MySQL \\ SO\end{tabular}}    & GPT 3.5            & 0.317            & 0.331             & \textbf{0.348}                 \\ \cline{2-5} 
                                                                        & GPT 4              & 0.365            & 0.407             & \textbf{0.441}                 \\ \hline
\multirow{2}{*}{\begin{tabular}[c]{@{}c@{}}MySQL \\ Forum\end{tabular}} & GPT 3.5            & \textbf{0.395}   & 0.387             & 0.382                          \\ \cline{2-5} 
                                                                        & GPT 4              & 0.400            & 0.360             & \textbf{0.449}                 \\ \hline
\multirow{2}{*}{\begin{tabular}[c]{@{}c@{}}PG \\ SO\end{tabular}}       & GPT 3.5            & 0.256            & 0.290             & \textbf{0.321}                 \\ \cline{2-5} 
                                                                        & GPT 4              & 0.286            & 0.338             & \textbf{0.398}                 \\ \hline
\end{tabular}
}
\end{table}

\stitle{Exp-3: How does document representation model work in document retrieval?}
%
We compare \textbf{\sys-docs-only} with document representation methods: \textbf{(1) Q-only}: only retrieving most similar questions using Sentence-BERT. \textbf{(2) M-only}: only retrieving most similar manuals using Sentence-BERT. \textbf{(3) No train} : retrieving similar questions and manuals using Sentence-BERT.

\begin{table*}
\caption{Document retrieval recall of different documents representation methods in the NL evaluation setting.}
\label{tab:doc_retrieval_recall}
 \resizebox{\textwidth}{!}
 {
 \begin{tabular}{|c|cccc|cccc|cccc|}
\hline
\textbf{Datasets}     & \multicolumn{4}{c|}{\textbf{MySQL SO}}                                                                                          & \multicolumn{4}{c|}{\textbf{MySQL Forum}}                                                                                        & \multicolumn{4}{c|}{\textbf{PG SO}}                                                                                              \\ \hline
\textbf{top-$k$}        & \multicolumn{1}{c|}{\textbf{1}}     & \multicolumn{1}{c|}{\textbf{3}}     & \multicolumn{1}{c|}{\textbf{5}}    & \textbf{10}    & \multicolumn{1}{c|}{\textbf{1}}     & \multicolumn{1}{c|}{\textbf{3}}     & \multicolumn{1}{c|}{\textbf{5}}     & \textbf{10}    & \multicolumn{1}{c|}{\textbf{1}}     & \multicolumn{1}{c|}{\textbf{3}}     & \multicolumn{1}{c|}{\textbf{5}}     & \textbf{10}    \\ \hline \hline
Q-only                & \multicolumn{1}{c|}{0.265}          & \multicolumn{1}{c|}{0.36}           & \multicolumn{1}{c|}{0.412}         & 0.496          & \multicolumn{1}{c|}{0.257}          & \multicolumn{1}{c|}{0.461}          & \multicolumn{1}{c|}{0.494}          & 0.544          & \multicolumn{1}{c|}{0.203}          & \multicolumn{1}{c|}{0.277}          & \multicolumn{1}{c|}{0.315}          & 0.357          \\ \hline
M-only                & \multicolumn{1}{c|}{0.218}          & \multicolumn{1}{c|}{0.345}          & \multicolumn{1}{c|}{0.391}         & 0.45           & \multicolumn{1}{c|}{0.253}          & \multicolumn{1}{c|}{0.368}          & \multicolumn{1}{c|}{0.392}          & 0.458          & \multicolumn{1}{c|}{0.158}          & \multicolumn{1}{c|}{0.175}          & \multicolumn{1}{c|}{0.228}          & 0.381          \\ \hline
No Train              & \multicolumn{1}{c|}{0.298}          & \multicolumn{1}{c|}{0.453}          & \multicolumn{1}{c|}{0.533}         & 0.615          & \multicolumn{1}{c|}{0.233}          & \multicolumn{1}{c|}{0.445}          & \multicolumn{1}{c|}{0.526}          & 0.581          & \multicolumn{1}{c|}{0.203}          & \multicolumn{1}{c|}{0.277}          & \multicolumn{1}{c|}{0.312}          & 0.438          \\ \hline
\sys-docs-only & \multicolumn{1}{c|}{\textbf{0.327}} & \multicolumn{1}{c|}{\textbf{0.461}} & \multicolumn{1}{c|}{\textbf{0.54}} & \textbf{0.621} & \multicolumn{1}{c|}{\textbf{0.336}} & \multicolumn{1}{c|}{\textbf{0.492}} & \multicolumn{1}{c|}{\textbf{0.595}} & \textbf{0.672} & \multicolumn{1}{c|}{\textbf{0.299}} & \multicolumn{1}{c|}{\textbf{0.426}} & \multicolumn{1}{c|}{\textbf{0.473}} & \textbf{0.533} \\ \hline
\end{tabular}
}
\end{table*}

\begin{table}
\caption{F1-scores of knob diagnosis of different document representation methods (And-d:\sys-doscs-only).}
\label{tab:doc_retrieval_reasoning}
\resizebox{0.7\columnwidth}{!}{
\begin{tabular}{|c|c|c|c|c|c|}
\hline
\textbf{Datasets}                                                       & \textbf{Reasoners} & \textbf{Q-only} & \textbf{M-only} & \textbf{No Train} & \textbf{And-d} \\ \hline \hline
\multirow{2}{*}{\begin{tabular}[c]{@{}c@{}}MySQL \\ SO\end{tabular}}    & GPT 3.5            & 0.334           & 0.304           & 0.338             & \textbf{0.348}        \\ \cline{2-6} 
                                                                        & GPT 4              & 0.304           & 0.102           & 0.400             & \textbf{0.441}        \\ \hline
\multirow{2}{*}{\begin{tabular}[c]{@{}c@{}}MySQL \\ Forum\end{tabular}} & GPT 3.5            & 0.361           & 0.233           & 0.367             & \textbf{0.382}        \\ \cline{2-6} 
                                                                        & GPT 4              & 0.220           & 0.04            & 0.381             & \textbf{0.449}        \\ \hline
\multirow{2}{*}{\begin{tabular}[c]{@{}c@{}}PG \\ SO\end{tabular}}       & GPT 3.5            & 0.298           & 0.244           & 0.251             & \textbf{0.321}        \\ \cline{2-6} 
                                                                        & GPT 4              & 0.308           & 0.03            & 0.261             & \textbf{0.398}        \\ \hline
\end{tabular}
}
\end{table}

The results are reported in Table~\ref{tab:doc_retrieval_recall} and Table~\ref{tab:doc_retrieval_reasoning}. 
For document retrieval recall shown in Table~\ref{tab:doc_retrieval_recall}, \sys-docs-only outperforms the other three methods. Compared with Q-only and M-only, No Train performs better, showing the benefit of retrieving documents from multiple sources. However, No Train has limited performance compared with \sys-docs-only, which validates our claim that the semantics of documents from different sources may be heterogeneous. These results also demonstrate the effectiveness of our proposed contrastive learning approach that aligns documents from different sources into a unified representation space. 
%
%
Table~\ref{tab:doc_retrieval_reasoning} shows the F1-score of DBMS configuration debugging results with the top-$5$ retrieved documents. \sys-docs-only also outperforms other methods in all datasets, showing the benefits of the unified representation learning model.

\revised{\etitle{Different knob frequencies}. Next, we provide an in-depth analysis to examine the effect of knob frequency in historical questions. We first rank knobs according to their occurrence frequency in historical questions and then categorize them into high, medium, and low-frequency buckets.
}

\begin{figure}

    \includegraphics[width=0.99\columnwidth]{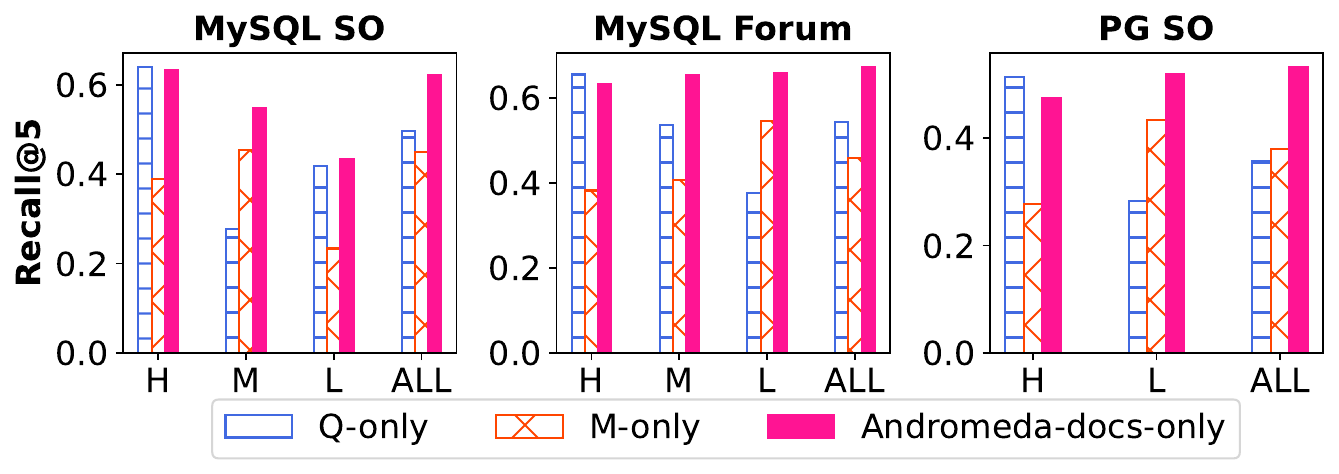}
    \caption{
    Document retrieval recall@5 in the NL evaluation setting by varying knob frequencies , where `H'/`M'/`L' means high/median/low frequency, respectively.
    }
    \label{fig:freq_retrieval}

\end{figure}

\revised{
We report the recall@5 in Figure~\ref{fig:freq_retrieval} for retrieval.
The results show that, during the retrieval stage, as the knob frequency decreases, the document retrieval recall of Q-only continuously decreases. This indicates that, if the knobs are rare in the historical question set, it is difficult to find useful information solely by retrieving similar questions. In contrast, the document retrieval recall of M-only performs similarly across different knob frequencies, and is lower than that of similar questions in high frequency. This suggests that due to the distribution differences between the questions and the manuals, finding similar manuals is a more challenging process. However, it is not affected by the frequency of the knob and is relatively effective for rare knobs. The performance of \sys-docs-only is generally superior
demonstrating that it is beneficial to retrieve documents from multiple sources and our contrastive learning method is effective.}
%

\begin{figure}
    \subfigure[Using GPT-3.5 for reasoning.]{
        \includegraphics[width=0.99\columnwidth]
        {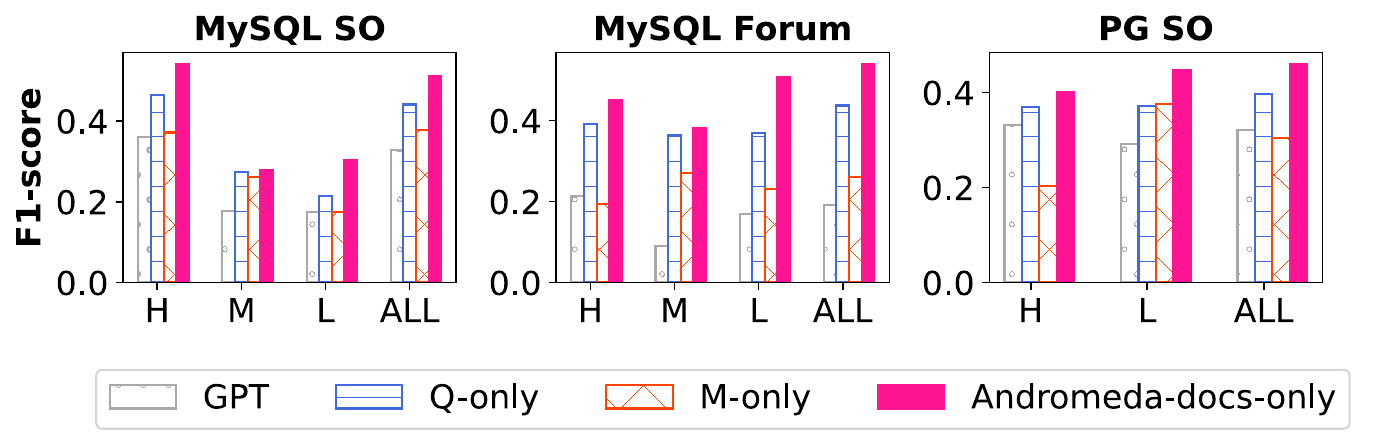}
    }
    \subfigure[Using GPT-4 for reasoning.]{
    \includegraphics[width=0.99\columnwidth]
        {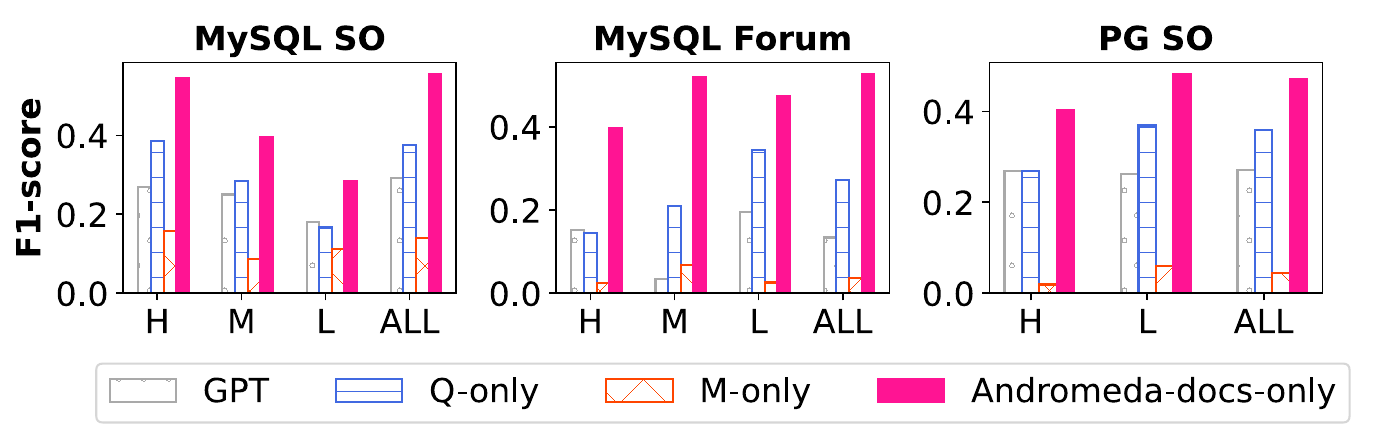}
    }
    \caption{F1-scores of knob diagnosis in the NL evaluation setting by varying knob frequencies, where `H'/`M'/`L' means high/median/low frequency, respectively.}
    \label{fig:freq_reasoning}
\end{figure}

\revised{
We report the F1-score in Figure~\ref{fig:freq_reasoning} for knob diagnosis, which illustrates the impact of different retrieval data sources on the knob diagnosis of LLMs under knobs of varying frequencies. For both GPT-3.5 and GPT-4, the F1-score of \sys-docs-only is generally higher than that of Q-only and M-only. Furthermore, with \sys-docs-only, the frequency of knobs has a smaller effect on the knob diagnosis of LLMs. This is attributed to the advantages of document representation with two types: question retrieval aids in high-frequency knobs, while manual retrieval compensates for the deficiencies of question retrieval in low-frequency knobs.}

\revised{\stitle{Finding 3: Combining different types of documents is more effective than using a single type of document. Also, our representation model can effectively align different types of documents, achieving better document retrieval results. }}

\begin{figure}
    \includegraphics[width=0.99\columnwidth]{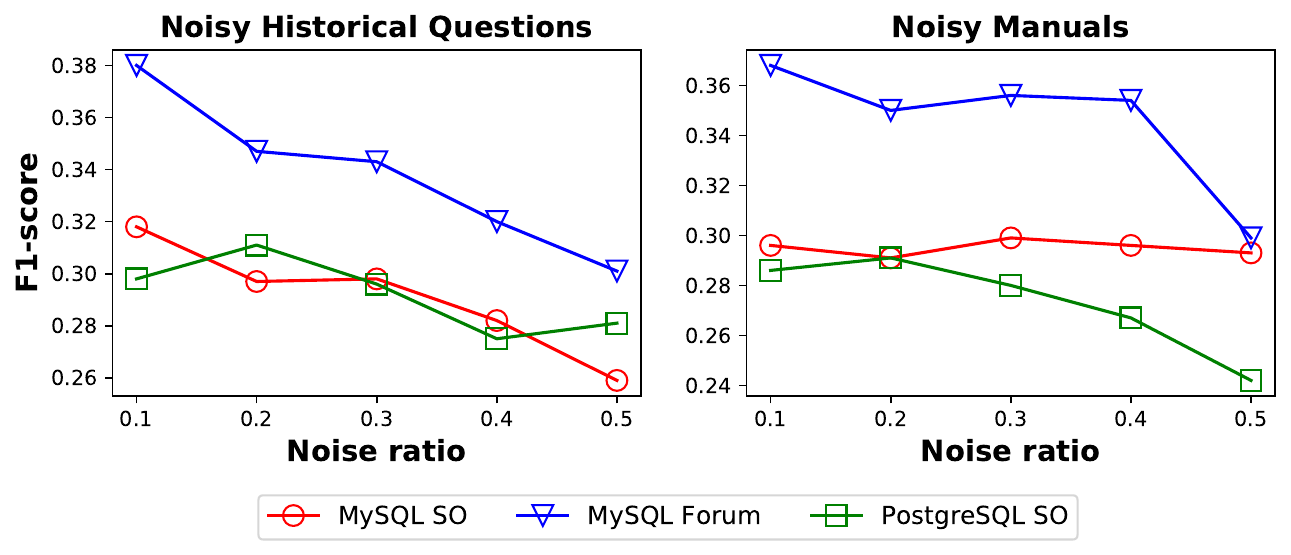}
    \caption{
   \revised{F1-score of knob diagnosis for retrieval document with different quality of source.}
    }
    \label{fig:noise_data}
\end{figure}

\revised{\stitle{Exp-4: How do inaccuracies in the retrieval sources impact the performance of \sys?} 
We manually inject noise into these sources by replacing the ground-truth labels in historical questions and database manuals with incorrect ones in various ratios, \ie 0.1, 0.2, 0.3, 0.4, and 0.5.}
%
%
\revised{Figure~\ref{fig:noise_data} shows that as the noise ratio increases, the overall performance of \sys decreases, highlighting the importance of the quality of retrieval sources. In particular, on some datasets (\eg MySQL SO in the Noisy Manuals setting), the performance does not significantly change. This is because, despite the presence of noisy documents, useful ones can still be retrieved to guide the LLMs.
} 
	

\revised{\stitle{Finding 4: The inaccuracies in the retrieval sources would impact the overall performance of \sys.
		%
	}}

\subsection{Evaluation on Telemetry Data Analysis} \label{sec:exp-tel}

\stitle{Exp-5: How does telemetry data analysis module work?}
To examine the effect of our telemetry data module, we compare GPTs and \sys with/without telemetry data analysis. We compare the quality of \textbf{(1) GPTs}, \textbf{(2) \sys-telemetry-only}: \sys without document retrieval,  \textbf{(3) \sys-docs-only}: \sys without telemetry data analysis, and \textbf{(4) \sys} respectively. 
%


\begin{table}
\caption{SuccessRates 
w/o telemetry analysis (And-d: Andromeda-docs-only; And-t: Andromeda-telemetry-only).}
\label{tab:telemetry}
 \resizebox{0.7\columnwidth}{!}{
\begin{tabular}{|c|c|c|c|c|c|}
\hline
\textbf{Datasets}                 & \textbf{Reasoners} & \textbf{GPTs}   & \textbf{And-t} & \textbf{And-d} & \textbf{\sys} \\ \hline \hline
\multirow{2}{*}{MySQL Run} & GPT 3.5   & 0.37 & 0.34 & 0.58 & \textbf{0.79}               \\ \cline{2-6} 
                           & GPT 4   & 0.43 & 0.64 & 0.63 & \textbf{0.76}               \\ \hline
\end{tabular}
}
\end{table}

The results are presented in Table~\ref{tab:telemetry}. GPTs and \sys-telemetry-only perform similarly. This indicates that using telemetry data analysis alone has limit effects as it only identifies a part of configuration issues that are highly related to telemetries: the telemetry data analysis can provide description on anomalous telemetry changes, which are useful for diagnosing anomalies caused by these anomalous changes. However, some questions are not due to the state of runtime environment resources, and thus using only telemetry data analysis is insufficient.

Comparing the results of \sys-docs-only and \sys, we can observe how telemetry data analysis works after adding document retrieval. The results show that \sys outperforms \sys-docs-only, demonstrating the effectiveness of telemetry data analysis. Question and manual retrieval can add more domain knowledge, while telemetry data analysis can further determine issues through metric changes. 

\begin{figure}
    \includegraphics[width=0.99\columnwidth]{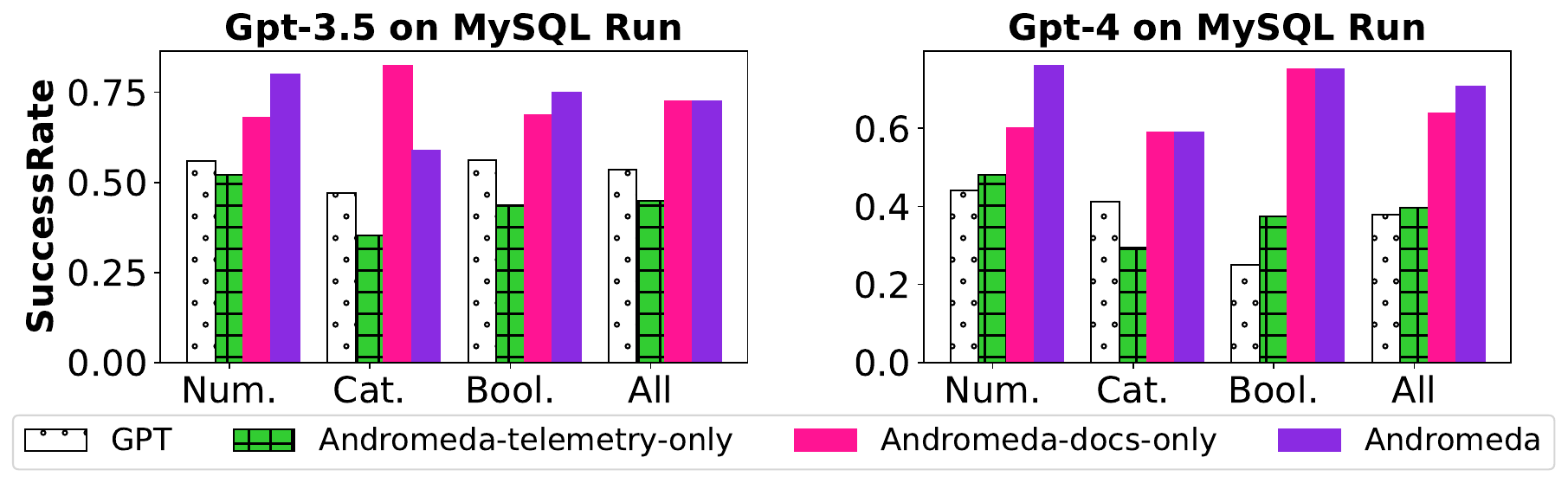}
    \caption{SuccessRates 
    on various knob types.}
    \label{fig:type}
\end{figure}

%
\revised{\etitle{Different knob type.} 
We provide an in-depth analysis to examine the effect of varying types of knobs by categorizing them based on their value ranges. According to the descriptions of knobs in the manual, we classify all knobs into Numeric, Categorical, and Boolean types. We calculate the accuracy to analyze the knobs based on these different categories in the Runnable setting. The results are reported in Figure~\ref{fig:type}. \sys outperforms other methods across various types of knobs, indicating that both telemetry data analysis and text retrieval are beneficial for knob diagnosis. More specifically, the impact of document retrieval is more pronounced, as \sys-docs-only generally performs better than \sys-telemetry-only. This is because related questions and manuals contain richer knowledge. However, for numeric knobs, we observe that the improvement of \sys-docs-only is not as significant for GPT-3.5 compared to other types of knobs. Moreover, \sys significantly outperforms \sys-docs-only on numeric knobs.  Numeric knobs typically control resource allocation and usage, such as \texttt{innodb\_buffer\_pool\_size} in MySQL, and these knobs often address issues caused by improper resource allocation. These issues are reflected in telemetry data analysis, making telemetry data analysis more helpful for diagnosing numeric knobs.}

\revised{\stitle{Finding 5: The telemetry data analysis module is helpful for DBMS configuration debugging. When including retrieved documents, the improvement from adding telemetry data analysis is more significant.}}

\subsection{Efficiency Evaluation} \label{sec:exp-eff}

\stitle{Exp-6: How efficient is \sys?}
We report the average runtime breakdown for diagnosis debugging in \sys, as shown in Table~\ref{tab:latency}. We have the following observations. First, the average end-to-end runtime
used to recommend knobs is in seconds, which means that \sys is efficient to meet DBMS configuration debugging requirements. Second, we find that telemetry data analysis and reasoning
take most of the time. For telemetry data analysis, all metrics need seasonal-trend decomposition and statistical test, which would take time. For reasoning, the LLM has a large number of parameters and thus requires some time to generate outputs. 


\begin{table}
\caption{Time of \sys (in seconds).}
\label{tab:latency}
 \resizebox{0.65\columnwidth}{!}{
\begin{tabular}{|c|c|c|c||c|}
\hline 
\textbf{\begin{tabular}[c]{@{}c@{}}Time\\ \end{tabular}} & \textbf{\begin{tabular}[c]{@{}c@{}}Document\\  Retrieval\end{tabular}} & \textbf{\begin{tabular}[c]{@{}c@{}}Telemetry \\ Analysis\end{tabular}} & \textbf{Reasoning} & \textbf{Total} \\ \hline \hline
GPT 3.5                                                          & 0.07                                                                   & 2.85                                                                   & 2.45               & 5.37           \\ \hline
GPT 4                                                            & 0.07                                                                   & 2.85                                                                   & 3.87               & 6.79           \\ \hline
\end{tabular}
}
\end{table}

\stitle{Finding 6: \sys can efficiently support automatic DBMS configuration debugging by answering a wide range of NL questions regarding configuration issues.}


\revised{
\stitle{Exp-7: Whether \sys effectively reduces the end-to-end time?} 
We evaluate whether \sys effectively reduces the end-to-end time (\ie operational burdens) for DBAs.
To this end, we employ two experienced DBAs to solve questions in the Runnable setting and provide them with historical questions, database manuals, and telemetry data. We then consider three strategies: (1) \term{using manuals}: DBAs manually search DBMS manuals to provide diagnostic suggestions for tuning knobs; (2) \term{using search engines}: DBAs use search engine to provide diagnostic suggestions for tuning knobs; (3) \term{using \sys}: DBAs follow \sys's recommendations to address the issues. We compare the average time taken by the DBAs to troubleshoot configuration issues, as well as the SuccessRate scores. 
As shown in Table~\ref{tab:user_study}, \sys achieves a higher SuccessRate than the alternative strategies and significantly reduces end-to-end time. These results demonstrate that \sys effectively reduces end-to-end time for DBAs while providing accurate troubleshooting suggestions.
}

\begin{table}
\caption{\revised{Comparison on SuccessRate and end-to-end time of using \sys with using manuals and using search engines in Runnable setting.}}
 \resizebox{0.65\columnwidth}{!}{
  \revised{
\begin{tabular}{|c|cc|cc|}
\hline
\textbf{Metrics}          & \multicolumn{2}{c|}{\textbf{SuccessRate}}     & \multicolumn{2}{c|}{\textbf{Time (minutes)}}       \\ \hline
\textbf{DBA Experts}      & \multicolumn{1}{c|}{\textbf{\# 1}} & \textbf{\# 2} & \multicolumn{1}{c|}{\textbf{\# 1}} & \textbf{\# 2} \\ \hline\hline
\term{using manuals}             & \multicolumn{1}{c|}{0.42}          & 0.38          & \multicolumn{1}{c|}{14.79}         & 24.51         \\ \hline
\term{using search engine}       & \multicolumn{1}{c|}{0.54}          & 0.46          & \multicolumn{1}{c|}{14.55}         & 9.75          \\ \hline
\term{using \sys} & \multicolumn{1}{c|}{\textbf{0.76}} & \textbf{0.76} & \multicolumn{1}{c|}{\textbf{7.55}} & \textbf{3.79} \\ \hline
\end{tabular}
}
\label{tab:user_study}
}
\end{table}

\revised{\stitle{Finding 7: \sys significantly reduces the end-to-end time for DBAs during database configuration debugging.
		%
	}}

\begin{figure}
    \includegraphics[width=0.99\columnwidth]{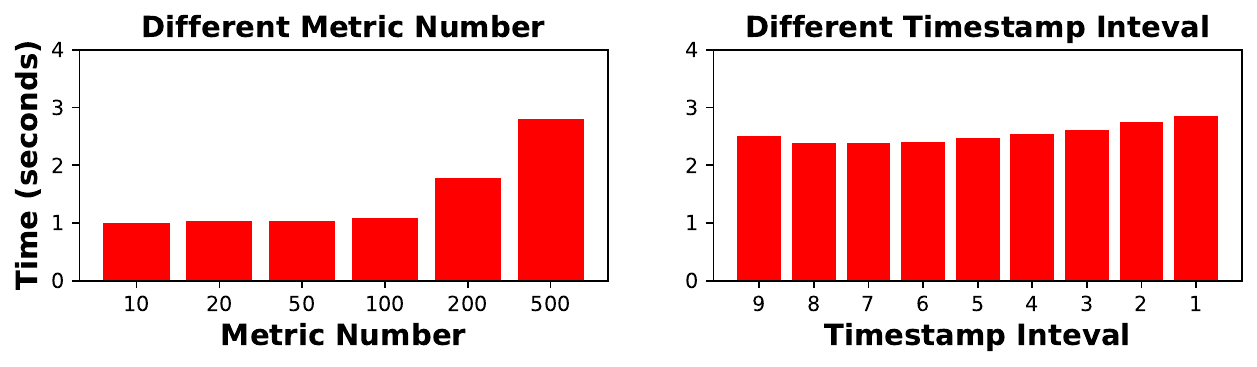}
    \caption{\revised{Evaluation on scalability of telemetry analysis.}}
    \label{fig:telemetry_scale}
\end{figure}

\revised{\stitle{Exp-8: How does \sys perform in terms of the scalability for telemetry data analysis.		
}
We vary the timestamp interval (\ie the sampling interval for each telemetry metric $s_{i}$) and the number of metrics (\ie the total count of telemetry metrics $s$).}
%
%
%
\revised{As shown in Figure~\ref{fig:telemetry_scale}, 
with the increase of metric number and the decrease of timestamp interval, the time required for telemetry analysis increases in a sub-linear manner. Moreover, the overall time remains within a few seconds, which is totally acceptable in our configuration debugging settings.
}
 

\revised{\stitle{Finding 8: 
		\sys performs well in terms of scalability for telemetry data analysis.
}}

\revised{\stitle{Exp-9: Whether \sys has good performance on cost efficiency?
} We respectively calculate the average cost regarding calling APIs for the approaches, namely LLMs, LLMs (all knobs) and \sys for GPT 3.5 and GPT 4. 
For each NL debugging question, we have  the results:
$0.0017$\$ for GPT-3.5, $0.053$\$ for GPT-4, $0.012$\$ for GPT-3.5 (all knobs), $0.039$\$ for GPT-4 (all knobs), $0.006$\$ for \sys-3.5 and $0.022$\$ for \sys-4. }

\revised{\stitle{Finding 9: \sys performs well on cost efficiency.}}

\section{Conclusions and Future Work} \label{sec:conclusion}

In this paper, we have introduced a novel framework \sys which serves as a natural surrogate of DBAs to answer a wide range of natural language (NL) questions regarding DBMS configuration issues, and generates recommendations to fix these issues.
\sys utilizes a retrieval-augmented generation (RAG) strategy that effectively provides domain-specific context for questions from multiple sources,
which significantly improves the performance of DBMS configuration debugging. 
Experiments on real-world DBMS configuration debugging datasets show that \sys significantly outperforms existing solutions.

\revised{
Several questions still remain to be answered.
First, besides DBMS configuration debugging, we can consider addressing more types of issues, such as SQL rewriting, index tuning, etc. Second, a more effective method should be proposed to assist the LLM in value recommendation for knobs. This is a more challenging problem because the range of values can be continuous.}

\begin{acks}
    We thank the anonymous reviewers for their constructive feedback. This work was partly supported by the NSF of China (62436010 and 62441230), Alibaba Innovative Research Program, 
    the Beijing Natural Science Foundation (L222006), the Research Funds of Renmin University of China, the Outstanding Innovative Talents Cultivation Funded Programs 2023 of Renmin University of China and Guangzhou Municipality Big Data Intelligence Key Lab (2023A03J0012).
\end{acks}

\newpage

\bibliographystyle{ACM-Reference-Format}
\bibliography{param-recommend}

\clearpage
\newpage
\appendix
\section{Case Study} \label{sec:case_study}

\revised{We present two cases to demonstrate how \sys works and identify areas for potential improvement.}



\vspace{-1em}
\subsection{\revised{Good Cases}}

\begin{figure*}
    \subfigure[\revised{Prompt}]{
    \includegraphics[width=0.99\textwidth]
        {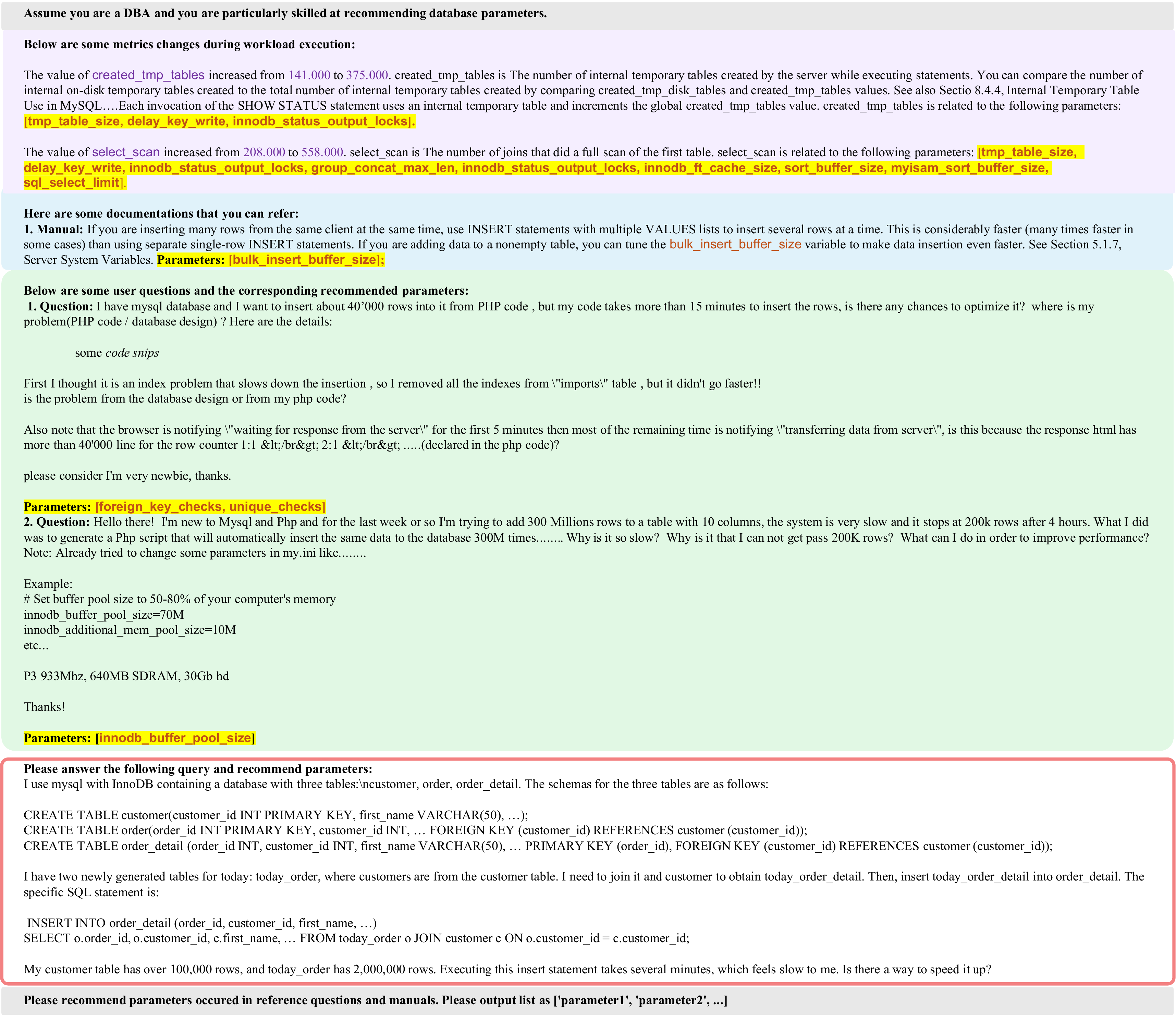}
    }
    \label{fig:casestudy_goodcase}
    \subfigure[\revised{Response}]{
    \includegraphics[width=0.45\textwidth]
        {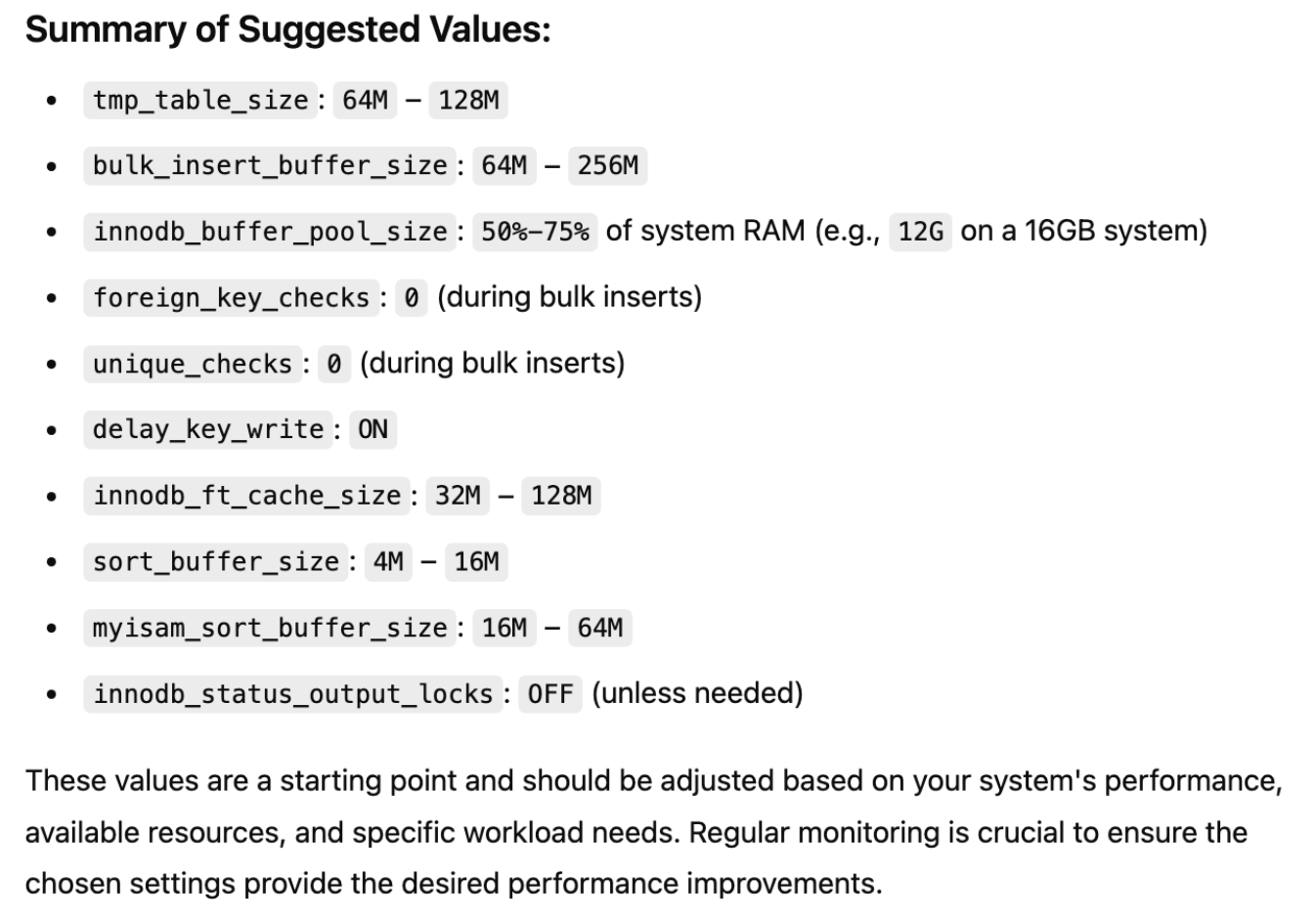}
    }
    
    \vspace{-1em}
    \caption{\revised{A good case that user encounters a slow insertion problem, and RAG with manuals, historical questions, and telemetry can effectively resolve the issue.}}
    \label{fig:good_response}
    \vspace{-1em}
\end{figure*}

\revised{The first case, shown in Figure~\ref{fig:good_response}(a), is similar to the running example in Figure~\ref{fig:running-example}.} 


\revised{\stitle{Prompt.} In the instruction phase, the first block (gray) represents the task, assigning the role and specific responsibilities to the LLM. As indicated in the user's question, represented by the fifth block (white), the user is working with MySQL using the InnoDB engine and needs to join two large tables for data insertion. However, the process takes several minutes, and thus the user seeks ways to optimize performance. 
Now, we discuss how \sys works.	

(1) For telemetry analysis, the second block (purple) provides details about changes, explaining two metrics: \texttt{created\_tmp\_tables} and \texttt{select\_scan}. It also recommends tuning related knobs, such as \texttt{tmp\_table\_size}, \texttt{delay\_key\_write}, etc. 

(2) The retrieved manual in the third block (blue) explains that using the INSERT command with multiple VALUES lists and adjusting the \texttt{bulk\_insert\_buffer\_size} knob can accelerate the insertion process, specifically recommending tuning the \texttt{bulk\_insert\_buffer\_size}. 

(3) The retrieved historical questions in the fourth block (green) highlight two related questions about slow insertion. The first involves slow insertion of approximately 40,000 rows into a database via PHP code, with recommendations to adjust \texttt{foreign\_key\_check} and \texttt{unique\_checks}. The second question concerns adding 300 million rows to a table at a slow pace, recommending the tuning of \texttt{innodb\_buffer\_pool\_size}.}

\revised{\stitle{Response.} For the response, the LLM predicts 10 configuration knobs along with their corresponding values. The \texttt{bulk\_insert\_buffer\_size} is recommended based on the manuals, while \texttt{foreign\_key\_check}, \texttt{unique\_checks}, and \texttt{innodb\_buffer\_pool\_size} are suggested by historical questions. The remaining knobs are derived from the telemetry analysis. We tested tuning each knob to the recommended values and found the following adjustments to be particularly effective:}

\revised{\begin{itemize}
    \item \texttt{tmp\_table\_size} from telemetry analysis.
    \item \texttt{bulk\_insert\_buffer\_size} from manual retrieval.
    \item \texttt{innodb\_buffer\_pool\_size} from historical questions.
    \item \texttt{foreign\_key\_checks} from historical questions.
    \item \texttt{unique\_checks} from historical questions.
    \item \texttt{sort\_buffer\_size} from telemetry analysis.
\end{itemize}}

\revised{Some knob values were predicted as a range in the response. We tested multiple values and found that more than one setting can be effective. For example, setting \texttt{tmp\_table\_size} to both $70M$ and $80M$ yielded positive results, indicating that various values within the range can work well.}

\revised{\stitle{Discussion.} With rich knowledge incorporated, the LLM accurately recommends the appropriate tuning knobs. For these multiple effective knobs, tuning a combination of them further accelerates insertion performance compared to adjusting a single knob. These knobs are sourced from various references, including manuals, historical questions, and telemetry data, highlighting the advantage of retrieval-augmented generation (RAG) from multiple sources.}

\subsection{\revised{Bad Cases}}

\revised{The second case illustrates a failure of \sys.}

\begin{figure*}
    \subfigure[\revised{Prompt}]{
        \includegraphics[width=0.90\textwidth]
        {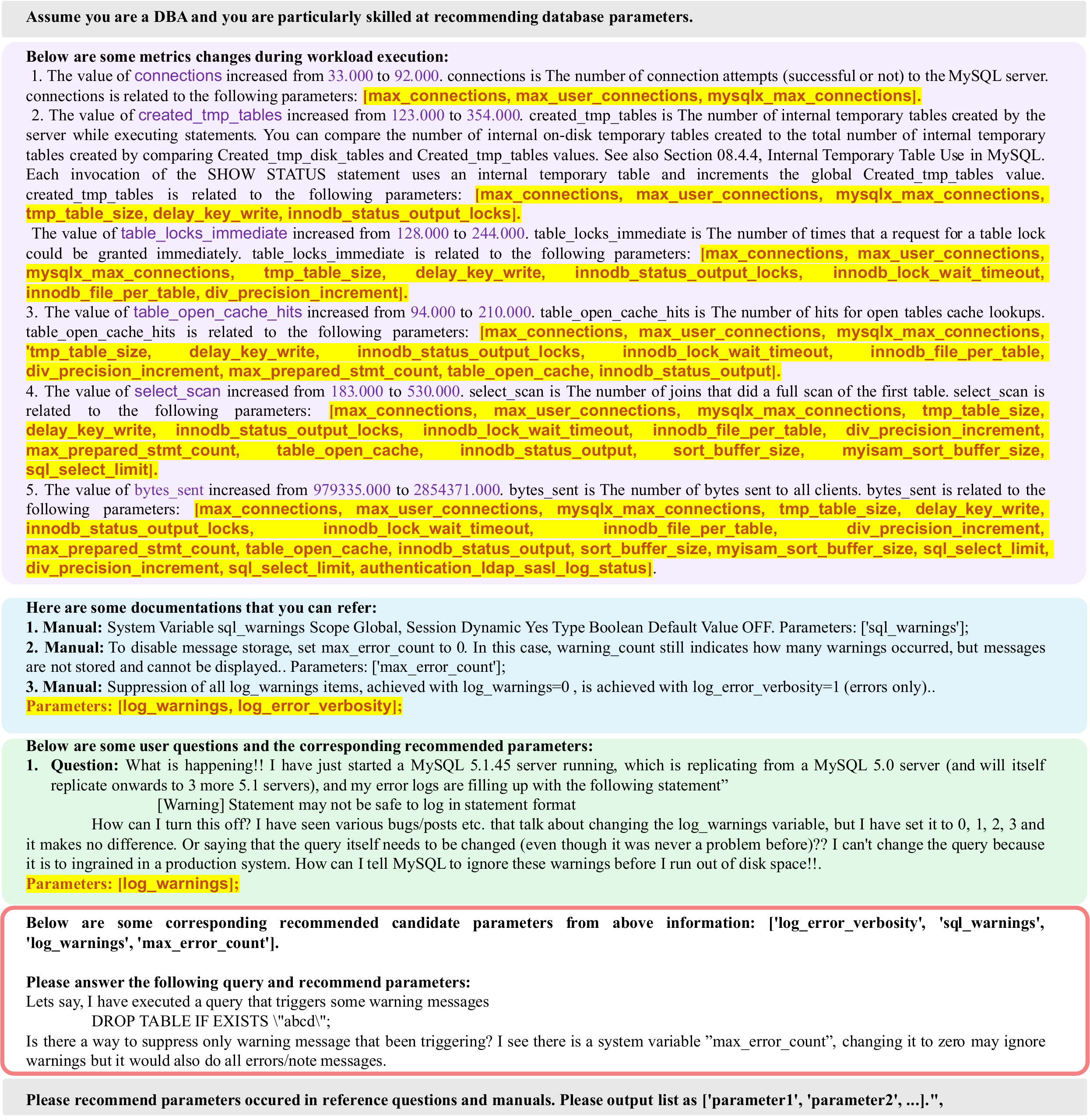}
    }
    \subfigure[\revised{Response}]{
    \includegraphics[width=0.4\textwidth]
        {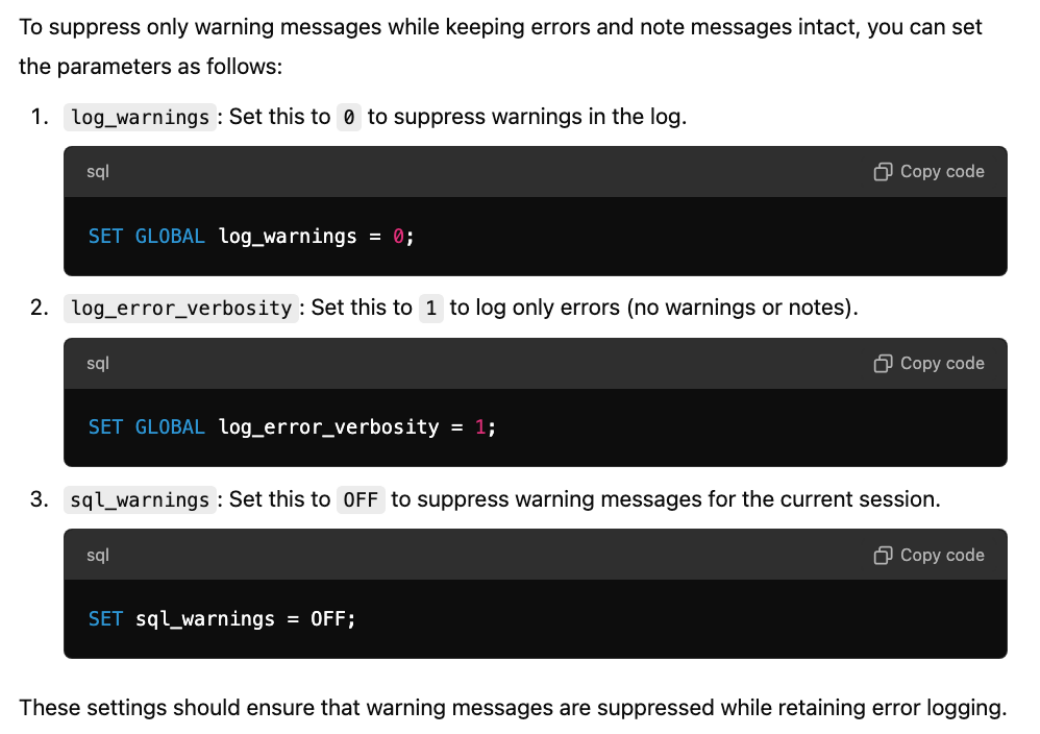}
    }

    \vspace{-1em}
    \caption{\revised{A bad case that user wants to suppress only the warning message triggered. \sys retrieves manuals of knobs with similar functions but can not directly solve the user's question.}}
    \label{fig:casestudy_badcase}
    \vspace{-1em}
\end{figure*}

\revised{\stitle{Prompt.} As shown in Figure~\ref{fig:casestudy_badcase}, the question (in the white block) shows that the user executed \texttt{DROP TABLE IF EXISTS "abcd"} and wanted to suppress only the warning message it triggered. Although he were aware of the \texttt{max\_error\_count} variable, he was concerned that setting it to zero would also suppress errors and note messages, which he wanted to avoid. 
	
(1) For the telemetry analysis (in the purple block), \sys reports changes and descriptions of five metrics: \texttt{connections}, \texttt{created\_tmp\_tables}, \texttt{table\_open\_cache\_hits}, \texttt{select\_scan}, and \texttt{bytes\_sent}, recommending adjustments to \texttt{max\_connections} and other related knobs. 

(2) In the manual retrieval (in the blue block), \sys retrieves three entries related to warning-related knobs. The first is a basic description of \texttt{sql\_warning}, the second covers the usage of \texttt{max\_error\_count}, and the last explains the usage of \texttt{log\_warnings} and \texttt{log\_error\_verbosity}. 

(3) For the historical questions (in the green block), \sys retrieves a question about how to suppress warnings in logs and recommended adjusting \texttt{log\_warnings}.}

\revised{\stitle{Response.} The LLM predicts \texttt{log\_warnings} based on both manuals and historical questions, while \texttt{sql\_warnings} and \texttt{log\_error\_verbosity} are suggested from the manuals.}

\revised{\stitle{Discussion.} The correct solution is to set \texttt{sql\_notes=0}, as detailed in the manual: the \texttt{sql\_notes} system variable controls whether note messages increment the \texttt{warning\_count} and whether the server stores them. By default, \texttt{sql\_notes} is set to 1, but when set to 0, notes do not increment the \texttt{warning\_count}, and the server does not store them. 
	
In this case, historical questions are not helpful, as the most relevant question deals with warnings in the log, not warning messages. Without related historical questions, this source proves unhelpful. Telemetry is also not useful since adjusting warning notes does not trigger any anomalies in telemetry. For the manuals, \sys retrieves several related entries for knobs with similar functions to \texttt{sql\_notes}, but none are directly relevant. The first is \texttt{sql\_warning}, which is irrelevant because it controls whether single-row \texttt{INSERT} statements produce an informational string when warnings occur (and this manual was not retrieved). The second is \texttt{max\_error\_count}, which the user has already tried. The third is \texttt{log\_warnings} and \texttt{log\_error\_verbosity}, which control warnings in logs but do not affect warning messages.}

\revised{\stitle{Improvement.} One of the limitations of the current version of \sys is its difficulty in distinguishing between similar knobs when retrieving documents. In this case, \texttt{sql\_notes}, \texttt{sql\_warning}, \texttt{max\_error\_count}, \texttt{log\_warnings}, and \texttt{log\_error\_verbosity} can all suppress warnings or errors, but in different contexts. For example, \texttt{log\_warnings} suppresses warnings in logs, whereas \texttt{sql\_notes} suppresses warning messages. To address this, we can improve the system by prompting LLMs to analyze the differences between the retrieved knobs and select the most relevant one based on the user's specific question and the retrieved documentation.}



\end{document}